\documentclass[a4paper,prb,twocolumn,showpacs,amsmath,amssymb]{revtex4}
\usepackage[english]{babel}
\usepackage[latin1]{inputenc}
\usepackage{amscd,amsmath,amssymb,verbatim}
\usepackage[dvips]{graphicx,color}
\usepackage{subfigure}
\usepackage{textcomp}
\usepackage{calc}
\usepackage{bbm}
\usepackage[OT1,T1]{fontenc}
\usepackage{natbib}
\usepackage[verbose,hypertexnames=false,bookmarksopenlevel=1,filecolor=blue,
linkcolor=blue,citecolor=blue,pdfstartview=FitH,bookmarksopen,bookmarksnumbered,
colorlinks,plainpages=false,linktocpage,ps2pdf]{hyperref}

\DeclareMathAlphabet{\mathpzc}{OT1}{pzc}{m}{it}

\DeclareSymbolFont{mathbold}{OML}{cmm}{b}{it}
\DeclareMathSymbol{\balpha}{\mathord}{mathbold}{11}
\DeclareMathSymbol{\bbeta}{\mathord}{mathbold}{12}
\DeclareMathSymbol{\bgamma}{\mathord}{mathbold}{13}
\DeclareMathSymbol{\bdelta}{\mathord}{mathbold}{14}
\DeclareMathSymbol{\bepsilon}{\mathord}{mathbold}{15}
\DeclareMathSymbol{\bvarepsilon}{\mathord}{mathbold}{34}
\DeclareMathSymbol{\bzeta}{\mathord}{mathbold}{16}
\DeclareMathSymbol{\bEta}{\mathord}{mathbold}{17}
\DeclareMathSymbol{\btheta}{\mathord}{mathbold}{18}
\DeclareMathSymbol{\bvartheta}{\mathord}{mathbold}{35}
\DeclareMathSymbol{\biota}{\mathord}{mathbold}{19}
\DeclareMathSymbol{\bkappa}{\mathord}{mathbold}{20}
\DeclareMathSymbol{\blambda}{\mathord}{mathbold}{21}
\DeclareMathSymbol{\bmu}{\mathord}{mathbold}{22}
\DeclareMathSymbol{\bnu}{\mathord}{mathbold}{23}
\DeclareMathSymbol{\bxi}{\mathord}{mathbold}{24}
\DeclareMathSymbol{\bpi}{\mathord}{mathbold}{25}
\DeclareMathSymbol{\bvarpi}{\mathord}{mathbold}{36}
\DeclareMathSymbol{\brho}{\mathord}{mathbold}{26}
\DeclareMathSymbol{\bvarrho}{\mathord}{mathbold}{37}
\DeclareMathSymbol{\bsigma}{\mathord}{mathbold}{27}
\DeclareMathSymbol{\bvarsigma}{\mathord}{mathbold}{38}
\DeclareMathSymbol{\btau}{\mathord}{mathbold}{28}
\DeclareMathSymbol{\bupsilon}{\mathord}{mathbold}{29}
\DeclareMathSymbol{\bphi}{\mathord}{mathbold}{30}
\DeclareMathSymbol{\bvarphi}{\mathord}{mathbold}{39}
\DeclareMathSymbol{\bchi}{\mathord}{mathbold}{31}
\DeclareMathSymbol{\bpsi}{\mathord}{mathbold}{32}
\DeclareMathSymbol{\bomega}{\mathord}{mathbold}{33}
\DeclareMathSymbol{\biGamma}{\mathord}{mathbold}{0}
\DeclareMathSymbol{\biDelta}{\mathord}{mathbold}{1}
\DeclareMathSymbol{\biTheta}{\mathord}{mathbold}{2}
\DeclareMathSymbol{\biLambda}{\mathord}{mathbold}{3}
\DeclareMathSymbol{\biXi}{\mathord}{mathbold}{4}
\DeclareMathSymbol{\biPi}{\mathord}{mathbold}{5}
\DeclareMathSymbol{\biSigma}{\mathord}{mathbold}{6}
\DeclareMathSymbol{\biUpsilon}{\mathord}{mathbold}{7}
\DeclareMathSymbol{\biPhi}{\mathord}{mathbold}{8}
\DeclareMathSymbol{\biPsi}{\mathord}{mathbold}{9}
\DeclareMathSymbol{\biOmega}{\mathord}{mathbold}{10}

\newcommand{\bra}[1]{\langle #1|}
\newcommand{\ket}[1]{|#1\rangle}
\newcommand{\braket}[2]{\langle #1|#2\rangle}

\newcommand{\goodgap}{\hspace{\subfigtopskip}\hspace{\subfigbottomskip}}

\newcommand{\kt}[1]{\text{\tiny{#1}}}

\newcommand{\feyn}[2]{\raisebox{-0.4\height}{\includegraphics[scale=#1]{#2}}}

\newcommand{\I}{\mathit{i}}

\begin{document}
\title{
Dimensional dependence of weak localization corrections and spin relaxation in quantum wires with Rashba spin-orbit coupling  }

\author{P. Wenk}
\email[]{p.wenk@jacobs-university.de}
\homepage[]{www.physnet.uni-hamburg.de/hp/pwenk/}
\affiliation{School of Engineering and Science, Jacobs University Bremen, Bremen 28759, Germany}
\author{S. Kettemann}
\email[]{s.kettemann@jacobs-university.de}
\homepage[]{www.physnet.uni-hamburg.de/hp/kettemann/}
\affiliation{School of Engineering and Science, Jacobs University Bremen, Bremen 28759, Germany and Asia Pacific Center for Theoretical
Physics and Division of Advanced Materials Science, Pohang University of Science and Technology (POSTECH), San31, Hyoja-dong, Nam-gu, Pohang 790-784, South Korea}

\begin{abstract}
The quantum correction to the conductivity in disordered quantum wires with linear Rashba spin-orbit coupling is obtained.  For  quantum wires with spin-conserving boundary conditions, we find a crossover from weak antilocalization to weak localization as the wire width $W$ is reduced using exact diagonalization of the Cooperon equation. This crossover is due  to the dimensional dependence of the spin relaxation  rate of conduction electrons, which becomes diminished, when the wire width $W$ is smaller than the bulk spin precession length $L_{SO}$. We thus confirm previous results for small wire width, $W/L_{SO}\lesssim 1$ [S. Kettemann, Phys. Rev. Lett.\textbf{98}, 176808(2007)], where only the transverse 0-modes of the Cooperon equation had been taken into account. We find  that spin helix solutions become stable for arbitrary ratios of linear Rashba and Dresselhaus coupling in narrow wires. For wider wires, the spin relaxation rate is found to be not monotonous as function of wire width $W$: it becomes first enhanced for $W$ on the order of the bulk spin precession length $L_{SO}$ before it becomes diminished for smaller wire widths. In addition, we find that the spin relaxation is smallest at the edge of the wire for wide wires. The effect of the Zeeman coupling  to the magnetic field perpendicular to the 2D electron system (2DES) is studied and found  to result in a modification of the magnetoconductivity: it shifts the crossover from weak antilocalization to weak localization to larger wire widths $W_c$. When the transverse confinement potential of the quantum wire is smooth, the boundary conditions become rather adiabatic. Then, the spin relaxation rate is found to be enhanced as the wire width $W$ is reduced. We find that only a spin polarized state retains a finite spin relaxation rate in such narrow wires. Thus, we conclude that the injection of polarized spins into nonmagnetic quantum wires should be favorable in wires with smooth confinement potential. Finally, in wires with tubular shape, corresponding to transverse periodic boundary conditions, we find  no reduction of the spin relaxation rate.
\end{abstract}
\pacs{ 72.10.Fk, 72.15.Rn, 73.20.Fz}
\maketitle
\section{Introduction}
Spintronic devices which rely on coherent spin precession of conduction electrons\cite{dasdatta,spintronics} require a small spin relaxation rate. As the electron  momentum is randomized due to disorder, spin-orbit (SO) interaction is expected to result not only in a spin precession but in randomization of the electron spin, the  D'yakonov-Perel'  spin relaxation with rate $ 1/\tau_s$.\cite{perel}
This spin relaxation is expected to vanish in narrow wires whose width $W$ is of the order of  Fermi wavelength $\lambda_F$,\cite{kiselev,meyer} since the back scattering from impurities can  in one-dimensional wires only  reverse  the SO field and thereby the spin precession. In this paper, we show, however, that  $1/\tau_s$ is already strongly reduced in much wider wires: as soon as the wire  width $W$ is smaller than bulk spin precession  length  $L_{\kt{SO}}$, which is the length on which the electron spin precesses a full cycle. This  explains the reduction of the spin relaxation rate in quantum wires for widths exceeding both the elastic mean-free path $l_e$ and  $\lambda_F$, as recently observed with  optical\cite{holleitner} as well as with  weak localization measurements.\cite{hu05_1,hu05_2,hu05_3,hu05_4,kunihashi:226601} Since $L_\kt{SO}$  can be several $\mu m$ and is not changed significantly as the wire width $W$ is reduced, the reduction of spin relaxation can be very useful for applications:
the spin of conduction electrons precesses coherently as it moves along the wire on  length scale $L_\kt{SO}$.
It becomes randomized and relaxes  on the longer length scale $L_s (W) = \sqrt{D_e \tau_s}$ only [$D_e= v_F l_e /2$ ($v_F$, Fermi velocity) is the 2D diffusion constant]. Quantum interference of electrons in low-dimensional, disordered conductors is known to result in corrections to the  electrical conductivity $\Delta \sigma$. This quantum correction, the weak localization effect, is a very sensitive tool to study dephasing and symmetry-breaking mechanisms in conductors.\cite{review_1,review_3,review_4} The entanglement of spin and charge by SO interaction reverses the effect of weak localization and thereby enhances the conductivity. This weak antilocalization effect was predicted by Hikami \textit{et al.}\cite{nagaoka} for conductors with  impurities of heavy elements. As conduction electrons scatter from such impurities, the SO interaction randomizes their spin. The resulting spin relaxation suppresses interference in spin triplet configurations. Since the time-reversal operation changes not only the sign of  momentum but also the sign  of the spin, the interference in singlet configuration remains unaffected. Since singlet interference reduces the electron's
return probability, it enhances the conductivity, which is named the weak antilocalization effect. In weak magnetic fields, the singlet
contributions are suppressed. Thereby, the conductivity is reduced  and the  magnetoconductivity becomes negative. The magnetoconductivity of wires is thus related to the magnitude of the spin relaxation rate. \\
In Sec. \ref{Sec:QuantTranspCor}, we first derive the quantum corrections to the conductivity for wires with general bulk SO interaction and relate it to the Cooperon propagator. In Sec. \ref{Sec:CoopSpinDiff}, we diagonalize the Cooperon for two-dimensional (2D) electron systems with Rashba SO interaction. We compare the spectrum of the triplet Cooperon with the one of the spin-diffusion equation. In Sec. \ref{Sec:SolutionCoopinWire}, we present the solution of the Cooperon equation for a wire geometry. We review the solutions of the spin-diffusion  equation in the wire geometry and compare the resulting spin relaxation rate  with the one extracted from the Cooperon equation. Then we proceed to calculate the quantum corrections to the conductivity using the exact diagonalization of the Cooperon propagator.
In the last part of this section, we consider two other kinds of boundary conditions. We calculate the spin relaxation rate in narrow wires with adiabatic boundaries, which arise in wires with smooth lateral confinement and regard also tubular wires. In Sec. \ref{Sec:MagnetoWithZeeman}, we study the influence of the Zeeman coupling to a magnetic field perpendicular to the quantum well in  a system with sharp boundaries and analyze how the magnetoconductivity is modified. In Sec. \ref{Sec:conclusions}, we draw the conclusions and compare with experimental results.
In Appendix \ref{derivationBoundary}, we give the derivation of the non-Abelian Neumann boundary conditions for the Cooperon propagator.
In Appendix \ref{RelaxTensor}, we show the connection between the effective vector potential $\bf A_S$ due to SO coupling and the spin relaxation tensor.
In Appendix \ref{weakLoc2D}, we give the exact quantum correction to the electrical conductivity in 2D.
In Appendix \ref{exactDiag},  we detail the diagonalization of the Cooperon propagator.
In the following, we set $\hbar=1$.

\section{Quantum Transport Corrections}\label{Sec:QuantTranspCor}
If the   host lattice of the electrons provides SO interaction, quantum corrections
to the conductivity have to be calculated in the basis of eigenstates of the Hamiltonian with
SO interaction
\begin{equation} \label{hamiltonian}
H = \frac{1}{2 m_e}   ({\bf p} +e {\bf A} )^2 +V({\bf x})
-\frac{1}{2} \gamma \bsigma\left({\bf B}+{\bf B}_\kt{SO}({\bf p})\right),
\end{equation}
where  $m_e$ is the effective electron mass.  
 ${\bf A}$ is the vector potential due to the external magnetic field 
  ${\bf B}$.  ${\bf B}_\kt{SO}^T = ({B_\kt{SO}}_x, {B_\kt{SO}}_y) $ is the momentum dependent SO field. $\bsigma$ is a vector, with components $\bsigma_i$, $i =x,y,z$,  the Pauli matrices, $\gamma$ is the gyromagnetic ratio with $\gamma=g \mu_\kt{B}$ with the effective g factor of the material, and $\mu_\kt{B} = e/2m_e$ is the Bohr magneton constant. For example,  the breaking of  inversion symmetry in 
   III-V semiconductors causes  a SO interaction, which for quantum wells grown in the $[ 001]$ direction is given by
\cite{dresselhaus}
\begin{equation}\label{B-Dresselhaus}
-\frac{1}{2} \gamma{\bf B}_\kt{SO,D} = \alpha_1 ( -\hat{e}_x p_x + \hat{e}_y p_y)
+ \gamma_D  ( \hat{e}_x p_x p_y^2 - \hat{e}_y p_y p_x^2).
\end{equation}
Here, $\alpha_1 = \gamma_D \langle p_z^2 \rangle$ is the linear Dresselhaus parameter, which measures the strength of the term  linear in    momenta $p_x, p_y$ in the plane of the 2DES. When  $\langle p_z^2 \rangle \sim 1/a^2 \ge k_F^2$  ($a$ is the thickness of the 2DES and $k_F$ is the Fermi wavenumber), that term  exceeds the cubic Dresselhaus terms which have  coupling strength $\gamma_D$. Asymmetric confinement of the 2DES yields  the  Rashba term which does not depend on the growth direction
\begin{equation}
-\frac{1}{2} \gamma{\bf B}_\kt{SO,R} = \alpha_2  ( \hat{e}_x p_y - \hat{e}_y p_x),
\end{equation}
 with $\alpha_2$ the  Rashba parameter.\cite{rashba_1,rashba_2} We consider the standard white-noise model for  the impurity potential, $V({\bf x})$, which vanishes on average
$\langle V ({\bf x}) \rangle=0$, is uncorrelated, $\langle V({\bf
x}) V({\bf x'})\rangle = \delta ({\bf x-x'})/2 \pi \nu \tau$, and
weak, $\epsilon_F\tau\gg 1$. Here, $\nu = m_e/(2 \pi)$ is the
average density of states per spin channel and $\tau$ is the elastic
scattering time. Going to momentum (${\bf Q}$) and frequency
($\omega$)  representation, and summing up  ladder diagrams, to take
into account  the  diffusive motion of the conduction electrons,
yield the quantum correction to the static conductivity as
\cite{nagaoka}
\begin{equation}\label{DeltaSigmaQ}
\Delta  \sigma = -2\frac{e^2}{2 \pi}\frac{D_e}{ \text{Vol.}}
\sum_{\bf Q} \sum_{\alpha, \beta = \pm} C_{\alpha \beta \beta
\alpha,  \omega=0} ({\bf Q}),
\end{equation}
where $\alpha,\beta = \pm$ are the spin indices, and the Cooperon propagator $\hat{C}$  is for $ \epsilon_F \tau \gg 1$ ($\epsilon_F$, Fermi energy), given by
\begin{eqnarray}
\hat{C}_{\omega = E - E^\prime}({\bf Q} =\mathbf{p}+\mathbf{p}^\prime)
= \tau
\left(1-\sum_{\mathbf{q}} \feyn{0.15}{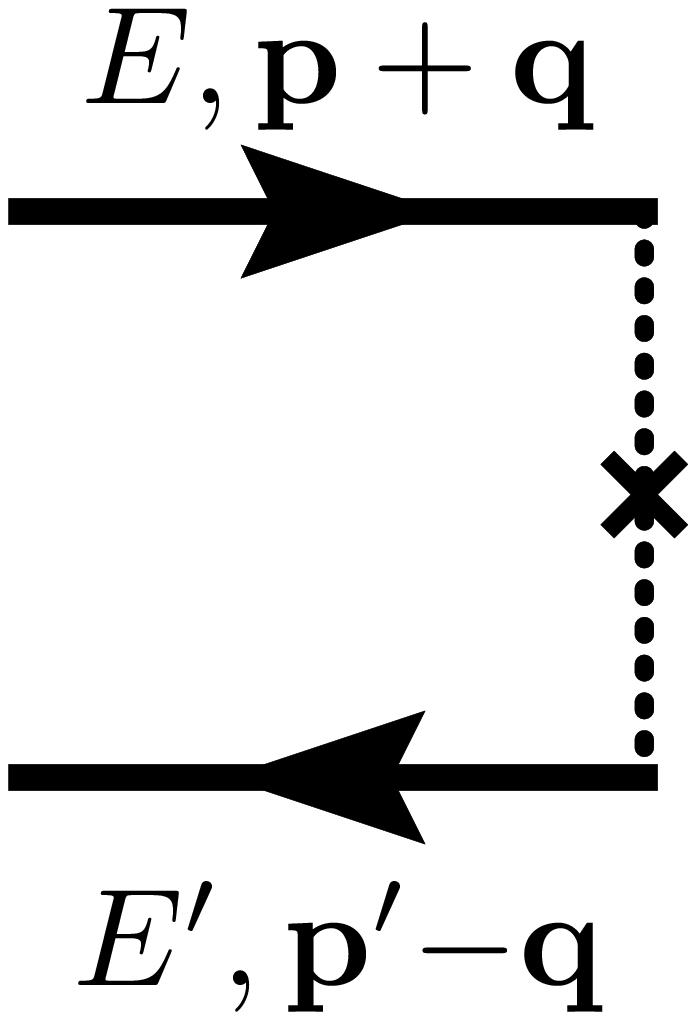}\enspace\right)^{-1}, \label{ladder}
\end{eqnarray}
where the impurity averaged electron propagator is given in the first Born approximation by
\begin{equation}
 \mathpzc{G}^{R}_E ({\bf p})=\includegraphics[scale=0.3]{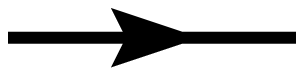}=\frac{1}{E-H_0({\bf p})+\I\frac{1}{2\tau}},
\end{equation}
and $ \mathpzc{G}^{A}_E ({\bf p})$ is its complex conjugate,
respectively.
$H_0$ is the Hamiltonian, Eq.\,(\ref{hamiltonian}), without disorder potential $V$.
The impurity vertex (the cross) is given by $1/2\pi \nu \tau$.
Impurity averaging products of Green's functions of the type
$\langle \mathpzc{G}^R\mathpzc{G}^R\rangle$ and $\langle
\mathpzc{G}^A\mathpzc{G}^A\rangle$ yield small corrections of order
$1/\epsilon_F \tau$. Thus, the problem reduces to the calculation
of the correlation function
\begin{align}
\sum_\mathbf{q}\feyn{0.15}{gamma1PartTextFINAL.eps}=&\frac{1}{2\pi\nu\tau}\sum_\mathbf{q}\mathpzc{G}_{E,\sigma}^{R}(\mathbf{p}+\mathbf{q})\mathpzc{G}_{E^\prime,\sigma^\prime}^{A}(\mathbf{p}^\prime-\mathbf{q}),
\intertext{which simplifies for weak disorder $\epsilon_{F} \tau \gg 1$ to}
 \approx & \int\frac{d\Omega}{2\pi}\frac{1}{1-\I\tau \hat{\Sigma}},
\end{align}
where
\begin{equation}
\hat{\Sigma}=\epsilon_{\mathbf{p}^\prime + \mathbf{q},\bsigma^\prime}-\epsilon_{\mathbf{p}-\mathbf{q},\bsigma}.\label{Sigma}
\end{equation}
For diffusive wires, for which the elastic mean-free path $l_e$ is
smaller than the wire width $W$, the integral is over all angles of
velocity $ {\bf v}$ on the Fermi surface. Using
\begin{eqnarray}
\epsilon_\mathbf{p}&=&\frac{(\mathbf{p}+e\mathbf{A})^2}{2m_e}-\frac{1}{2} \gamma \bsigma ({\bf B}+{\bf B}_\kt{SO}({\bf p})),\nonumber\\
\mathbf{v}&=&\frac{\mathbf{p-q}+e\mathbf{A}}{m_e},\nonumber\\
\mathbf{S}&=&\frac{1}{2}(\bsigma+\bsigma^\prime),\nonumber\\
\mathbf{Q}&=&\mathbf{p+p^\prime} \nonumber,
\end{eqnarray}
we obtain to lowest order in $\mathbf{Q}$,
\begin{eqnarray}
\hat{\Sigma}&=&-\mathbf{v}(\mathbf{Q}+2e\mathbf{A}+2m_e\hat{a} \mathbf{S})+(\mathbf{Q}+2 e\mathbf{A})\hat{a} \bsigma^\prime \nonumber\\
&&+\frac{1}{2} \gamma (\bsigma^\prime-\bsigma) {\bf B}.
\end{eqnarray}
Here, the SO couplings are combined in  the matrix
\begin{equation} \label{a}
\hat{a}=   \left(   \begin{array}{cc}
-\alpha_1 + \gamma_D k_y^2& - \alpha_2 \\
\alpha_2 & \alpha_1 - \gamma_D k_x^2
\end{array} \right).
 \end{equation}
Thus, the Cooperon becomes
\begin{eqnarray}
&&\hat{C}({\bf Q})^{-1}=  \frac{1}{\tau} \left( 1 -
\int \frac{d \Omega}{2\pi} \right. \nonumber\\ &&  \left.
\frac{1}{1+\I\tau(\mathbf{v}(\mathbf{Q}+2e\mathbf{A}+2m_e\hat{a} \mathbf{S}) + H_{\sigma'}
+ H_Z  )}\right),
\end{eqnarray}
where $ H_{\sigma'}=-(\mathbf{Q}+2 e\mathbf{A})\hat{a} \bsigma^\prime$ and the Zeeman coupling to the external magnetic field yields
\begin{equation}
H_Z=- \frac{1}{2} \gamma (\bsigma^\prime-\bsigma)\mathbf{B}.\label{zeemanTerm}
\end{equation}
It follows that for weak disorder and without Zeeman coupling, the Cooperon depends only on the total momentum $\mathbf{Q}$ and the total spin $ {\bf S}$. Expanding the Cooperon to second order in $( {\bf Q} +2 e {\bf A} + 2m_e \hat{a} {\bf S} )$ and performing  the angular integral which  is for 2D diffusion (elastic mean-free path $l_e$  smaller than wire width $W$) continuous from $0$ to $2 \pi$ and yields
\begin{equation}
  \hat{C} ({\bf Q})= \frac{1}{D_e( {\bf Q} + 2 e {\bf A} + 2 e  {\bf A}_{\bf S})^2 + H_{\gamma_D}  }.\label{Cooperon1}
\end{equation}
The effective vector potential due to SO interaction, ${\bf
A}_{\bf S} = m_e  \hat{\alpha} {\bf S}/e$ ( where $\hat{\alpha} =
\langle\hat{a}\rangle$ denotes the matrix Eq.\,(\ref{a}), as averaged
over angle), couples to total spin vector $ {\bf S}$ whose components are four by four matrices. The cubic
Dresselhaus coupling is found to reduce the effect of the linear one
to     $\tilde\alpha_1:=\alpha_1 - m_e \gamma_D \epsilon_F/2$. Furthermore, it gives
rise to the spin relaxation term in Eq.\,(\ref{Cooperon1}),
\begin{equation} \label{hgamma}
H_{\gamma_D} =  D_e (m_e^2\epsilon_F \gamma_D)^2 (S_x^2 + S_y^2).
\end{equation}
In the representation of the singlet,  $\ket{S=0;m=0} = ( \ket{\uparrow \downarrow} - \ket{\downarrow \uparrow})/\sqrt{2}  \equiv \ket{\rightleftarrows}$ and triplet states $\ket{S=1;m=0} = ( \ket{\uparrow \downarrow} + \ket{\downarrow \uparrow})/\sqrt{2}  \equiv\ket{\rightrightarrows},\ket{S=1;m=1}\equiv\ket{\upuparrows},\ket{S=1;m=-1}\equiv\ket{\downdownarrows}$,
$\hat{C}$  decouples into a singlet and a triplet sector. Thus, the  quantum conductivity is a sum of singlet and triplet terms
\begin{eqnarray} \label{qmc}
\Delta\sigma & = &- 2 \frac{e^2}{2\pi}\frac{D_e}{ \text{Vol.}}
\sum_{\bf Q} \left(
- \frac{1}{D_e( {\bf Q} + 2 e {\bf A})^2} \right. \nonumber \\
&& \left.+\sum_{m=0,\pm1}  \langle S=1,m \mid \hat{C} ({\bf Q}) \mid S=1,m\rangle  \right).\enspace
\end{eqnarray}
With the cutoffs due to dephasing $1/\tau_\varphi$ and elastic scattering
$1/\tau$, we can integrate over all possible
wave vectors $\bf Q$ in the 2D case analytically (Appendix
\ref{weakLoc2D}).\\
In 2D, one can treat the magnetic field
nonperturbatively using  the basis of  Landau bands.\cite{nagaoka,knap,miller,af01,lg98,golub}
In wires with   widths smaller than  cyclotron length $k_F l_B^2$ ($l_B$, the magnetic length,  defined by $ B l_B^2 = 1/e$),  the Landau basis is not suitable. There is another way  to treat  magnetic fields:
quantum corrections are due to the interference between closed time-reversed paths. In magnetic fields, the electrons  acquire a magnetic phase, which breaks  time-reversal invariance.  Averaging over all  closed  paths,  one obtains a   rate with which the
magnetic field breaks the time-reversal invariance, $1/\tau_B$. Like  the dephasing rate $1/\tau_{\varphi}$, it cuts off
the divergence arising from  quantum corrections with  small wave vectors ${\bf Q}^2 < 1/ D_e\tau_B$. In 2D  systems, $\tau_B$ is the time an electron  diffuses along a closed path enclosing  one magnetic flux quantum, $ D_e\tau_B = l_B^2$.
In wires of finite width $W$ the area which the  electron path  encloses  in a time $\tau_B$ is   $ W \sqrt{D_e\tau_B} $. Requiring that this   encloses one flux quantum gives $1/\tau_B = D_e e^2 W^2 B^2/3$. For arbitrary  magnetic field, the relation 
\begin{equation} \label{taub}
 1/\tau_B= D_e(2 e)^2 B^2 \langle y^2 \rangle,
\end{equation}
with the  expectation value of the square of the transverse position $\langle y^2\rangle$, yields $1/\tau_B=\left(1-1/(1+W^2/3l_B^2)\right)D_e/l_B^2$. Thus,  it is sufficient to  diagonalize  the Cooperon propagator as given by  Eq.\,(\ref{Cooperon1}) without magnetic field, as we will do in  the next chapters, and to add   the magnetic rate $1/\tau_B$  together with  dephasing rate $1/\tau_{\varphi}$ to the denominator of $\hat{C} ({\bf Q})$ when calculating the conductivity correction,  Eq.\,(\ref{qmc}).
\section{The Cooperon and  Spin Diffusion in  2D}\label{Sec:CoopSpinDiff}
\begin{figure}[t]
 \begin{center}
   \subfigure[]{\scalebox{1.2}{\includegraphics[width=\columnwidth/2]{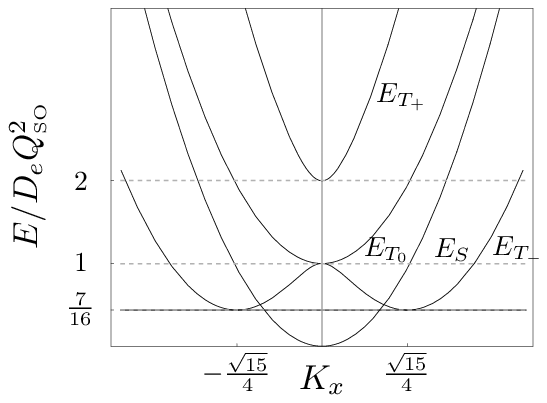}}}\goodgap
   \subfigure[]{\scalebox{0.7}{\includegraphics[width=\columnwidth/2]{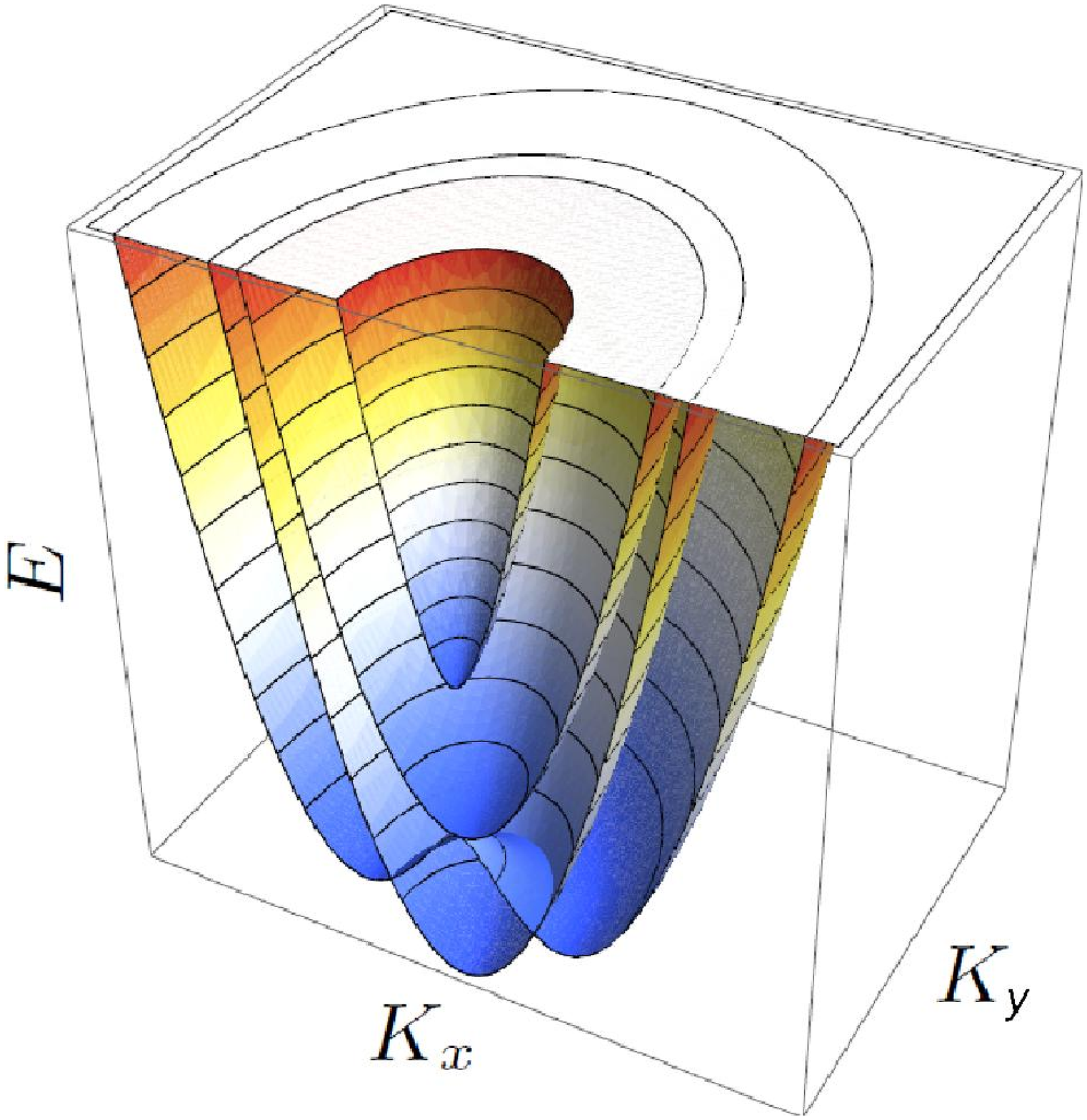}}}
   \caption{(Color online) 2D spectrum of $H_c$, $K_i=Q_i/Q_\kt{SO}$.}\label{plot:free_Hc}
  \end{center}
\end{figure}

\begin{figure}[htbp]
\begin{center}
 \includegraphics[width=\columnwidth]{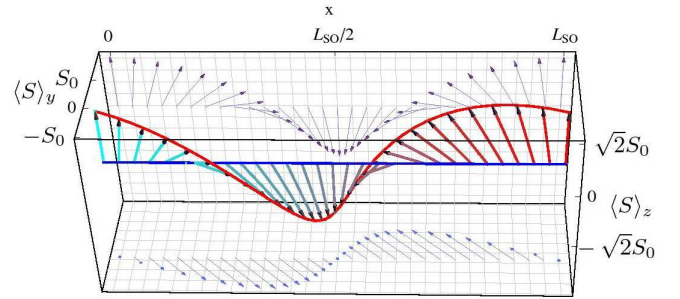}
\caption{(Color online) Persistent  spin helix solution of the spin-diffusion equation for 
  equal magnitude of linear Rashba and linear Dresselhaus coupling, Eq.\,(\ref{Eq:spinhelix}).
}\label{Figspinhelix}
\end{center}
\end{figure}

The Cooperon can be diagonalized analytically in 2D for pure Rashba coupling, $\alpha_1=0,\gamma_D=0$. For this case, we define the Cooperon Hamilton operator as 
\begin{equation}\label{free_Cooperon}
    H_c:=\frac{\hat{C}^{-1}}{D_e}={\bf Q}^2+2Q_\kt{SO}(Q_y S_x-Q_x S_y)+Q_\kt{SO}^2(S_y^2+S_x^2),
\end{equation}
with $Q_\kt{SO}=2 m_e \alpha_2 = 2\pi/L_\kt{SO},$ where $L_\kt{SO}$ is the   spin precession length.
In the representation of the singlet $\ket{\rightleftarrows}$ and triplet modes, $\{\ket{\upuparrows},\ket{\rightrightarrows},\ket{\downdownarrows}\}$
it becomes
\begin{equation}\label{free_H_c}
H_c= \left(
          \begin{array}{cccc}
{\bf Q}^2  & 0 & 0 & 0 \\
0 & Q_\kt{SO}^2+{\bf Q}^2 & \sqrt{2} Q_\kt{SO} Q_+ & 0 \\
0 &        \sqrt{2} Q_\kt{SO} Q_- & 2 Q_\kt{SO}^2+{\bf Q}^2 & \sqrt{2} Q_\kt{SO} Q_+ \\
0 &         0 & \sqrt{2} Q_\kt{SO} Q_- & Q_\kt{SO}^2+{\bf Q}^2 \\
\end{array}
\right),
\end{equation}
with $Q_\pm=Q_y\pm\I Q_x$.
  Diagonalization yields the gapless singlet eigenvalues and
the three triplet Cooperon eigenvalues
 with a gap due to the SO coupling (see Fig.\,\ref{plot:free_Hc}),
\begin{eqnarray}
    E_{S}(Q)/D_e &=& {\bf Q}^2 \label{eig_val_Coop2D_singlet},\\
  E_{T_0}(Q)/D_e &=& {\bf Q}^2+Q_\kt{SO}^2 \label{eig_val_Coop2D_T0},\\
E_{T_\pm}(Q)/D_e &=& {\bf Q}^2+\frac{3}{2}Q_\kt{SO}^2\pm\frac{Q_\kt{SO}^2}{2}\sqrt{1+16 \frac{ {\bf Q}^2}{Q_\kt{SO}^2}}, \label{eig_val_Coop2D_Tpm}
\end{eqnarray}
where  $E_S$  denotes the singlet eigenvalue and
 $E_{T_0}, E_{T_{\pm}}$ the three triplet eigenvalues. Notice that the two minima of the lowest triplet eigenmode are shifted to $Q=\pm(\sqrt{15}/4)  Q_\kt{SO}$ with a minimal eigenvalue of
$E/D_e=(7/16)Q_\kt{SO}^2$. As we show in the following,
this gap in the triplet modes
 is directly related to the D'yakonov-Perel'
 spin relaxation rate $1/\tau_s$.
We can get a better understanding of  the spin relaxation induced by the SO coupling and impurity
scattering by considering directly the spin-diffusion equation for  the expectation value of the electron-spin vector
\cite{MalshukovPRB.61.R2413}
\begin{equation}
 \mathbf{s}({\bf r},t) = \frac{1}{2} \langle \psi^\dagger({\bf r},t )\bsigma  \psi ({\bf r},t ) \rangle,
\end{equation}
where $\psi^\dagger = (\psi_+^\dagger,\psi_-^\dagger)$ is the
two-component vector of the up (+), and down (-) spin fermionic
creation operators and $\psi$ the two-component vector of annihilation
operators, respectively. In the presence of SO coupling, the
spin-diffusion equation  becomes for $v_F\mid\nabla_{\bf
r}\mathbf{s}\mid\ll 1/\tau$,
\begin{align}
0 ={}& \partial_t\mathbf{s}+\frac{1} {\hat{\tau}_s} {\bf s}-D_e{\bf \nabla}^2 {\bf s}\nonumber\\
&+ \gamma ({\bf B}-2 \tau  \langle ( \mathbf{\nabla}  {\bf v}_F) {\bf B}_{\rm SO} ({\bf {p}})
\rangle)
\times {\bf s}\quad\label{diffusioneq}\\
\intertext{and we define accordingly the spin-diffusion Hamiltonian $H_\kt{SD}$}
 0 ={}& \partial_t\mathbf{s}+D_e H_\kt{SD} \mathbf{s},
\end{align}
where the matrix elements of the spin relaxation terms are given by \cite{dyakonov72_2,dyakonov72_3}(Appendix \ref{RelaxTensor})
\begin{equation} \label{dp}
\frac{1}{\tau_{sij}} =  \tau\gamma^2\left( \langle  {\bf B}_{\rm SO}({\bf k})^2  \rangle \delta_{i j }-\langle { B}_{\rm SO}({\bf k})_i { B}_{\rm SO}({\bf k})_j\rangle\right).
\end{equation}
For pure Rashba SO interaction, the spin-diffusion operator
$H_\kt{SD}$ is in momentum representation\cite{Raimondi:2006}
\begin{equation}\label{Hsd}
H_\kt{SD}=\left(
\begin{matrix}
\frac{1}{D_e\tau_s}+{\bf k}^2 & 0 & -\I 2  Q_\kt{SO} k_x \\
0 & \frac{1}{D_e\tau_s}+{\bf k}^2 & -\I 2 Q_\kt{SO} k_y \\
\I 2 Q_\kt{SO}  k_x & \I 2 Q_\kt{SO}  k_y &
\frac{2}{D_e\tau_s}+{\bf k}^2
\end{matrix}
\right),
\end{equation}
with $ 1/D_e\tau_s = Q_\kt{SO}^2$. In the 2D case,
diagonalization yields  the eigenvalues
\begin{eqnarray}
 E_0(k)&=&{\bf k}^2+\frac{1}{D_e\tau_s},\label{SpinDiff2DE1}\\
 E_\pm(k)&=&{\bf k}^2+\frac{3}{2}\frac{1}{D_e\tau_s}\pm\frac{1}{2D_e\tau_s}\sqrt{1+16 \frac{ {\bf k}^2}{ Q_\kt{SO}^2 } }.\label{SpinDiff2DE23}
\end{eqnarray}
Thus, we find  that the spectrum of the spin-diffusion operator and the  one of the triplet
Cooperon Hamiltonian are identical in 2D (Ref. \onlinecite{PhysRevB.56.6436}) as long as time-reversal symmetry  is not broken. This confirms that antilocalization in the presence of  SO interaction, which has its cause in the suppression of the triplet modes in Eq.\,(\ref{qmc}),  is indeed  a direct measure of the spin relaxation.
  Mathematically, there exists a unitary transformation
\begin{equation}\label{UHSD}
  H_c=U_\kt{CD}H_\kt{SD}U_\kt{CD}^\dagger,
  \end{equation}
\begin{equation}\label{UHSDrepresent}
 U_\kt{CD}=
\left(
\begin{array}{ccc}
 -\frac{1}{\sqrt{2}} & \frac{\I}{\sqrt{2}} & 0 \\
 0 & 0 & 1 \\
 \frac{1}{\sqrt{2}} & \frac{\I}{\sqrt{2}} & 0
\end{array}
\right),
\end{equation}
with the according transformation between spin-density components $s_i$ and the triplet components of the Cooperon density $\tilde s$,
\begin{eqnarray}\label{spinDensityTriplet}
 \frac{1}{\sqrt{2}}(-s_x+\I s_y)  = \tilde s_\upuparrows,\\
 s_z=\tilde s_\rightrightarrows,\\
 \frac{1}{\sqrt{2}}(s_x+\I s_y)=\tilde s_\downdownarrows.
\end{eqnarray} 
This is a consequence of the fact that the four-component vector of charge density $\rho = (\rho_+ + \rho_-)/2$ and
spin-density vector ${\bf S}$ are related to the density vector ${\hat \rho}$ with the four  components $\langle \psi^\dagger_{\alpha} \psi^{\phantom{\dagger}}_{\beta} \rangle/\sqrt{2}$, where $\alpha, \beta = \pm$, by a unitary transformation. The classical evolution of the four-component density vector ${\hat \rho}$  is by definition governed by the diffusion operator, the diffuson.  The diffuson is related to the Cooperon in momentum space by substituting  ${\bf Q\rightarrow p-p^\prime}$ and the sum of the spins of the retarded and advanced parts, $\bsigma$ and $\bsigma^\prime$, by  their difference. Using this substitution, Eq.\,(\ref{free_Cooperon}) leads thus to the inverse of the diffuson propagator
\begin{equation}
    H_d:=\frac{\hat{D}^{-1}}{D_e}={\bf Q}^2+2Q_\kt{SO}(Q_y \widetilde{S}_x-Q_x \widetilde{S}_y)+Q_\kt{SO}^2(\widetilde{S}_y^2+\widetilde{S}_x^2),
\end{equation}
with $\widetilde{\bf S}=(\bsigma^\prime-\bsigma)/2$, which has the same spectrum as the Cooperon, as long as
the time-reversal symmetry  is not broken. In the representation of singlet and triplet modes the diffusion Hamiltonian becomes
\begin{equation}
H_d=\left(
\begin{matrix}
 2Q_\kt{SO}^2+{\bf Q}^2 & \sqrt{2}Q_\kt{SO}Q_-  & 0         & -\sqrt{2}Q_\kt{SO}Q_+\\
\sqrt{2}Q_\kt{SO}Q_+    & Q_\kt{SO}^2+{\bf Q}^2 & 0         & 0                    \\
0                       & 0                     & {\bf Q}^2 & 0                    \\
-\sqrt{2}Q_\kt{SO}Q_-   & 0                     & 0         & Q_\kt{SO}^2+{\bf Q}^2
\end{matrix}
\right). \label{representation:DiffusonWithSO}
\end{equation}
Comparing Eqs. (\ref{free_H_c}) and Eq.\,(\ref{representation:DiffusonWithSO}),  we  see that diagonalization leads to Eqs. (\ref{eig_val_Coop2D_singlet})-(\ref{eig_val_Coop2D_Tpm}).\\ 
It can be seen from Eqs. (\ref{SpinDiff2DE1}) and (\ref{SpinDiff2DE23}) that in the case of a homogeneous Rashba field, the spin-density has a finite decay rate. However, if we go beyond the pure Rashba system and include 
 a linear Dresselhaus coupling, the first term in Eq.\,(\ref{B-Dresselhaus}),  we can find  spin states which do not relax and are thus 
  persistent. The spin relaxation tensor, Eq.\,(\ref{dp}), acquires nondiagonal elements and changes to 
\begin{equation} \label{spinrelaxation}
         \frac{1}{\hat{\tau}_s} (k_F)= 4  \tau k_F^2  \left( \begin{array}{ccc}
           \frac{1}{2} \alpha^2 & - \alpha_1 \alpha_2  & 0  \\
         -   \alpha_1 \alpha_2 &  \frac{1}{2} \alpha^2 & 0 \\ 
               0 &  0 &  \alpha^2
         \end{array} \right), 
\end{equation}
with $\alpha=\sqrt{\alpha_1^2+\alpha_2^2}$. For $Q=0$ and $\alpha_1=\alpha_2=\alpha_0$, we find indeed a vanishing eigenvalue with a spin-density vector parallel to the spin-orbit field, ${\bf s}=s_0(1,1,0)^T$. Moreover there are two additional modes which do not decay in time but are inhomogeneous in space: the persistent spin helices,\cite{bernevig,liu:235322,PhysRevLett.83.4196,spinhelix,Koralek2009}
\begin{align}\label{Eq:spinhelix}
{\bf S} ={}& S_0  \left( \begin{array}{r}
           1 \\
          -1 \\ 
           0 
         \end{array} \right)  \sin \left(   \frac{2 \pi}{L_\kt{SO}} (x-y ) \right)\nonumber\\
         &+ S_0 \sqrt{2}  \left( \begin{array}{r}
           0 \\
           0 \\ 
           1 
         \end{array} \right)  \cos \left(   \frac{2 \pi}{L_\kt{SO}} (x-y ) \right),
\end{align} 
(Fig.\,\ref{Figspinhelix}) and the linearly independent solution, obtained by interchanging 
$\cos$ and $\sin$. Here,  $L_\kt{SO}=\pi/m_e \sqrt{2}\alpha_0$.
One has to keep in mind that this solution is not an eigenstate anymore in a quantum wire. However, we will show that there exist also long persisting solutions in a quasi-1D case.\\
It is worth to mention that in the case where cubic Dresselhaus coupling in Eq.\,(\ref{B-Dresselhaus}) cannot be neglected, the strength of linear Dresselhaus coupling $\alpha_1$ is shifted\cite{Kettemann:PRL98:2007} 
to $\tilde\alpha_1=\alpha_1 - m_e\gamma_D\epsilon_F/2$, as mentioned in Sec. \ref{Sec:QuantTranspCor}, and, e.g., in the $Q=0$ case, the spin relaxation rate becomes
\begin{align}
\frac{1}{\tau_s}={}& 2p_F^2\frac{\left(\alpha_2^2- \tilde\alpha_1^2\right)^2}{\alpha_2^2+ \tilde\alpha_1^2} \tau+D_e(m_e^2\epsilon_F\gamma_D)^2.
\end{align}
The condition for persistence is thus rather $\tilde\alpha_1=\alpha_2$. This  has been confirmed in a recent measurement (Ref. \onlinecite{Koralek2009}). The existence of such long-living modes has an effect on the quantum corrections to the  conductivity. In this case, $\tilde\alpha_1=\alpha_2=\alpha_0$, there is only weak localization in 2D.\cite{PhysRevB.51.16928,scheid:266401} In the next sections we will make use of the equivalence of the triplet sector of the  Cooperon propagator and the spin-diffusion propagator in quantum wires with appropriate  boundary conditions and show how  long-living modes may change the quantum corrections to the conductivity.
\section{Solution of the Cooperon Equation in Quantum Wires}\label{Sec:SolutionCoopinWire}
\subsection{Quantum Wires with Spin-Conserving Boundaries}
The conductivity of quantum wires with width $W < L_\varphi =\sqrt{D_e\tau_\varphi}$ is without SO interaction  dominated by  the transverse zero-mode $Q_y =0$. This yields  the quasi-1D weak localization correction.\cite{kurdak} However,  in the presence of SO interaction, setting simply $Q_y =0$ is not correct. If we consider spin-conserving boundaries, rather one has  to solve the Cooperon equation with the following modified boundary conditions as derived in Appendix \ref{derivationBoundary} (Refs. \onlinecite{af01,meyer}):
\begin{align} \label{bc}
\left(-\frac{\tau}{D_e}{\bf n}\cdot\langle{\bf v}_F [\gamma{\bf B}_\kt{SO}({\bf k})\cdot{\bf S}]\rangle-\I\partial_{\bf n}\right) C |_{\partial S} &= 0,\\
\intertext{where $\langle ... \rangle$ denotes the average over the direction of ${\bf v}_F$ and ${\bf k}$ which we rewrite using Eq.\,(\ref{AsBso}) for the given geometry as}
(-\I \partial_y + 2 e  {\bf (A_{S})}_y ) C\left( x, y = \pm
\frac{W}{2}\right) &= 0,\enspace\forall x,\quad\quad
\end{align}
where ${\bf n}$ is the unit vector normal to the boundary $\partial S$ and x is the coordinate along the wire. The transverse zero-mode $Q_y =0$ does not satisfy this condition. Therefore, it is convenient to  perform a non-Abelian gauge transformation,\cite{af01,MalshukovPRB.61.R2413} so that the transformed problem has Neumann  boundary conditions, and the transformed Cooperon Hamiltonian can therefore be diagonalized in zero-mode approximation for quantum wires. Since in quantum wires these boundary conditions apply only in the transverse direction, a transformation acting in the transverse  direction is needed:
$\hat{C}\rightarrow \tilde{\hat{C}} = U_A^{\phantom{\dagger}} \hat{C}U_A^\dagger$, with
$U_A = \exp (\I 2 e {\bf (A_S)}_y y)$. Then, the boundary condition
simplifies to $-\I \partial_y \tilde{C}(x,y=\pm
W/2)=0,\enspace\forall x$, and the Hamiltonian changes to
\begin{eqnarray}
    \tilde{H}_c&=&{\bf Q}^2-2Q_\kt{SO} Q_x [\cos(Q_\kt{SO} y)S_y-\sin(Q_\kt{SO} y)S_z ] \nonumber\\
               &&+Q_\kt{SO}^2 [\cos^2(Q_\kt{SO} y)S_y^2+\sin^2(Q_\kt{SO} y)S_z^2\nonumber\\
               &&-\sin(Q_\kt{SO} y)\cos(Q_\kt{SO} y)(S_y S_z+S_z S_y)] \\
               &=& ({\bf Q}  +2 e {\bf \tilde{A}}_s )^2. \label{transformed_Cooperon}
\end{eqnarray}
where   the effective vector potential 
${\bf A}_s$, as introduced in Eq.\,({\ref{Cooperon1}}), 
\begin{equation}
{\bf A}_s =  \frac{m_e}{e}\hat\alpha{\bf S}= \frac{m_e}{e}\left(
\begin{array}{ccc}
 0 & -\alpha_2 & 0 \\
 \alpha_2 & 0 & 0
\end{array}
\right)
\left(\begin{array}{c}
S_x\\
S_y\\
S_z
\end{array}\right),
\end{equation}
is transformed  to the effective vector potential $\tilde{\bf A}_s$ after the transformation $U_A$ has been applied to the Hamiltonian \begin{align}\label{A_s_transformed}
{\bf \tilde{A}}_s \equiv{}&\frac{m_e}{e}\tilde{\hat{\alpha}}(y){\bf S}\nonumber\\
={}& \frac{m_e}{e}\left(
\begin{array}{ccc}
 0 & -\alpha_2 \cos (Q_\kt{SO}y) & -\alpha_2 \sin (Q_\kt{SO}y) \\
 0 & 0 & 0
\end{array}\right)
\left(\begin{array}{c}
S_x\\
S_y\\
S_z
\end{array}\right),
\end{align}
which varies with the transverse coordinate $y$ on the length scale of
$L_\kt{SO}$. Now, we can see already that for narrow wires $W < L_\kt{SO}$, this vector potential varies linearly with $y$,  ${\bf \tilde{A}}_s \sim -m_e \alpha_2 Q_\kt{SO}y/e$, like the vector potential of the external magnetic field ${\bf B}$. Thus, it follows, that
for $W < L_\kt{SO}$, the spin relaxation rate is $1/\tau_s \sim Q_\kt{SO}^2 \langle y^2 \rangle \sim Q_\kt{SO}^2 W^2/12 $, vanishing  for small wire widths. As announced at the beginning, we thus see that the presence of boundaries  diminishes the spin relaxation already at wire widths of the order of $L_\kt{SO}$. If we include only pure Neumann boundaries to the Hamiltonian $H_c$, i.e., using the wrong covariant derivative, this would not affect the absolute spin relaxation minimum and it would be equal to the nonzero one in the 2D case.
We give a more precise answer in the following.
\subsection{Zero-Mode Approximation}\label{Zero-Mode}
\begin{figure}[htbp]
\begin{center}
 \includegraphics[width=3\columnwidth/4]{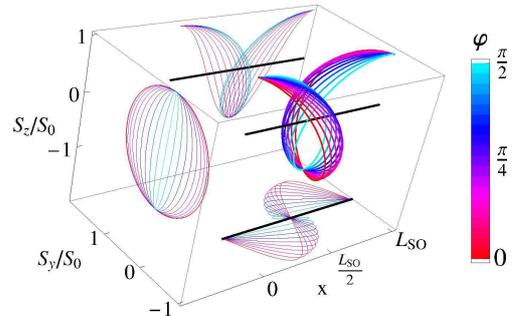}
\caption{(Color online) Long persisting spin helix solution of the spin-diffusion equation in a quantum wire whose width $W$ is smaller than the spin precession length $L_\kt{SO}$ for varying ratio of linear Rashba $\alpha_2 = \alpha \sin \varphi $ and linear Dresselhaus coupling, $\alpha_1 = \alpha \cos \varphi $, Eq.\,(\ref{Eq:spinhelix1d}), for  fixed $\alpha$ and $L_\kt{SO} = \pi/m_e \alpha$.}\label{spinhelix1d}
\end{center}
\end{figure}

For $W<L_{\varphi}$, we can  use  the fact that the  $n$th transverse nonzero-modes contribute terms to the conductivity which are by a factor $W/n L_{\varphi} $ smaller than  the 0-mode term, with $n$ a nonzero integer number. Therefore,  it should be a good approximation  to diagonalize the effective quasi-one-dimensional Cooperon propagator, which is  the transverse 0-mode expectation value of the   transformed inverse Cooperon propagator, Eq.\,(\ref{transformed_Cooperon}), $\tilde{H}_{1D} = \langle0\mid   \tilde{H}_c \mid 0\rangle$. It is crucial to note that $\tilde{H}_{1D}$  contains additional terms, created  by  the non-Abelian  transformation, which shows that taking just the transverse zero-mode approximation of the untransformed Eq.\,(\ref{free_Cooperon}) would yield a different, incorrect result. We can  now  diagonalize $\tilde{H}_{1D}$ and finally find the dispersion of quasi-1D triplet modes
\begin{eqnarray}\label{e0}
  && \frac{E_{t 0}}{D_e} =  Q_x^2 +  \frac{1}{2} Q_\kt{SO}^2    t_\kt{SO},
     \nonumber \\
  && \frac{E_{t \pm}}{D_e}  =   Q_x^2  +\frac{1}{4}  Q_\kt{SO}^2
     \nonumber \\
  && \times\left( 4 -  t_\kt{SO}
     \pm \sqrt{  t_\kt{SO}^2 +  64  \frac{ Q_x^2}{Q_\kt{SO}^2} (1+c_\kt{SO}(c_\kt{SO}-2) )}\right),\nonumber\\
\end{eqnarray}
where $ c_\kt{SO} $ and $ t_\kt{SO} $ are functions of the wire width $W$ as given by
\begin{equation}
  c_\kt{SO} = 1-  \frac{2 \sin (Q_\kt{SO} W/2)}{Q_\kt{SO} W},~
  t_\kt{SO} = 1-  \frac{ \sin (Q_\kt{SO} W)}{Q_\kt{SO} W}.
\end{equation}
\\
Inserting Eq.\,(\ref{e0}) into the expression for the quantum correction to the conductivity Eq.\,(\ref{qmc}), taking into account the magnetic field by inserting the magnetic rate $1/\tau_B(W)$ and the finite  temperature by inserting the dephasing rate $1/\tau_{\varphi}(T)$,
it remains to perform the integral over momentum $Q_x$,  as has been done in Ref. \onlinecite{Kettemann:PRL98:2007}.
For $Q_\kt{SO}W < 1$, the weak localization correction can then be written as 
\begin{eqnarray} \label{wl1D}
 && \Delta \sigma =\nonumber\\
 && \frac{\sqrt{H_W}}{\sqrt{H_{\varphi}+ B^*(W)/4}} - \frac{\sqrt{H_W}}{\sqrt{ H_{\varphi}
    + B^*(W)/4 +  H_{s}(W) }}
    \nonumber \\
 &&-2\frac{\sqrt{H_W}}{\sqrt{H_{\varphi}+ B^*(W)/4 + H_{s}(W)/2}}
\end{eqnarray}
in units of $e^2/2 \pi$.
We defined $H_W =1 /4 e W^2$ and  the effective external magnetic field
\begin{equation} \label{beff}
   B^*(W) =  \left(1-\left(1+\frac{W^2}{3 l_B^2}\right)^{-1}\right) B.
\end{equation}
The spin relaxation  field $H_{s}(W)$ is  for $Q_\kt{SO}W < 1$,
\begin{equation} \label{hso0}
 H_s (W) = \frac{1}{12}(Q_\kt{SO}W)^2  H_{s},
\end{equation}
suppressed in proportion to $(W/L_\kt{SO})^2$ similar to $B^*(W)$, Eq.\,(\ref{beff}).  Here, $H_s = 1/4 e D_e \tau_s$, with $1/\tau_{s} = 2 p_F^2  \alpha_2^2  \tau  $. As mentioned above, the analogy to the suppression of the effective magnetic field, Eq.\,(\ref{beff}), is expected, since the SO coupling enters  the transformed Cooperon, Eq.\,(\ref{transformed_Cooperon}), like an effective  magnetic vector potential.\cite{falko} Cubic Dresselhaus coupling, however,  would  give rise to an additional spin relaxation term, Eqs. (\ref{hgamma}) and (\ref{spinRelaxTensorVektorfeld}),  which has no  analogy to a magnetic field and is therefore not suppressed in diffusive wires although it is width dependent due to presence of modified Neumann boundaries. When $W$ is larger than SO length $L_\kt{SO}$, coupling to higher transverse modes may become relevant even if $W < L_{\varphi}$ is still satisfied, since the SO interaction may introduce coupling to higher transverse modes.\cite{aleiner} We will study these corrections by  numerical exact diagonalization in the next section. One  can expect that in  ballistic wires, $l_e>W$,  the spin relaxation rate is suppressed in analogy  to the flux cancellation effect, which yields the weaker rate, $1/ \tau_s (W)  = (W/C l_e) (D_e W^2/ 12
L_\kt{S}^4)$, where $C= 10.8$.\cite{km02_1,km02_2,km02_3}
Before we investigate the exact diagonalization in the pure Rashba case, we consider  an anisotropic field with linear Rashba and Dresselhaus SO coupling to see which form the long persisting spin-diffusion modes have in narrow wires. Also, here, we can take advantage of the equivalence of Cooperon and spin-diffusion equation as far as time-reversal symmetry is not violated.
We find three solutions whose spin relaxation rate decay proportional to $W^2$ for $\alpha_2\neq\alpha_1$ and which are persistent for $\alpha_2=\alpha_1$. The first solution is ${\bf s}=s_0(\alpha_1,\alpha_2,0)^T$ for $Q_x=0$ which is aligned with the effective SO field ${\bf B}_\kt{SO}({\bf k})=-2\gamma k_x(\alpha_1,\alpha_2,0)^T$. In this case, we have according to Eq.\,(\ref{hso0}) $H_\kt{s,RD1} (W) = (1/12)(\tilde Q_\kt{SO}W)^2 H_{s}$, with $\tilde Q^2_\kt{SO}=(2m_e(\alpha_1^2-\alpha_2^2))^2/\alpha^2$ and $1/\tau_{s} = 2 p_F^2\alpha^2\tau $, $\alpha=\sqrt{\alpha_1^2+\alpha_2^2}$. As mentioned above by transforming the vector potential $\bf A_S$, Eq.\,(\ref{A_s_transformed}), this alignment occurs due to the constraint on the spin-dynamics imposed by the  boundary condition as soon as the wire width $W$ is smaller  than the spin precession length $L_\kt{SO}$. In addition, we find two spin helix solutions in narrow wires,
\begin{equation}\label{Eq:spinhelix1d}
           {\bf s} = s_0  \left( \begin{array}{c}
           -\frac{\alpha_2}{\alpha} \\
            \frac{\alpha_1}{\alpha}  \\ 
               0 
         \end{array} \right)  \sin \left(   \frac{2 \pi}{L_\kt{SO}} x \right)
         + s_0   \left( \begin{array}{r}
          0 \\
          0  \\ 
          1 
         \end{array} \right)  \cos \left(   \frac{2 \pi}{L_\kt{SO}} x \right),
\end{equation}
and the linearly independent solution, obtained by interchanging $\cos$ and $\sin$ in Eq.\,(\ref{Eq:spinhelix1d}). The form of this long persisting spin helix depends therefore on the ratio of linear Rashba and linear Dresselhaus coupling strength, Fig.\,\ref{spinhelix1d}, and its spin relaxation rate is diminished as $H_\kt{s,RD2/3}=(1/2)H_\kt{s,RD1}$.
\subsection{Exact Diagonalization}
\begin{figure}[htbp]
 \begin{center}
    \includegraphics[width=\columnwidth]{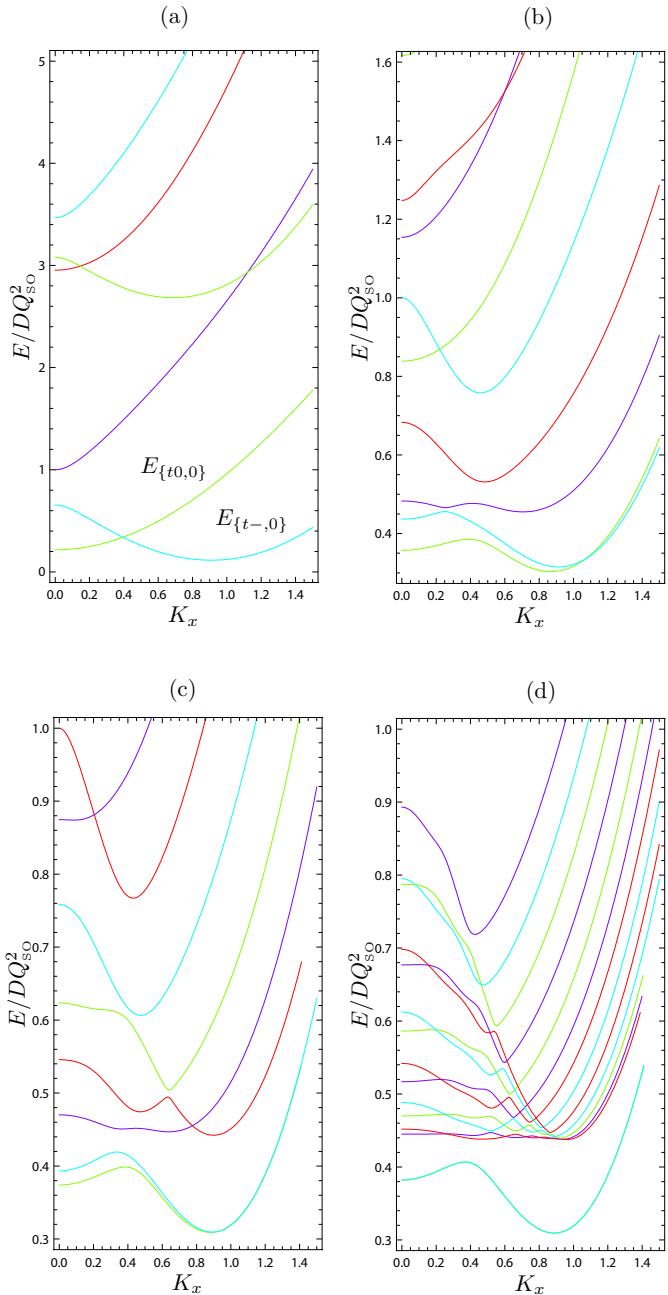}\\
    \caption{(Color online) Dispersion  of the triplet Cooperon modes for different dimensionless wire units $Q_\kt{SO}W$: (a) $Q_\kt{SO}W=2$, (b) $Q_\kt{SO}W=8$, (c) $Q_\kt{SO}W=12$, (d) $Q_\kt{SO}W=30$,
    plotted as function of $K_x = Q_x/Q_\kt{SO}$. For $Q_\kt{SO}W\gg 3$,  $E_{\{t0,0\}}$ and $E_{\{t-,0\}}$
    evolve into degenerate   branches for large $K_x$.  (For $Q_\kt{SO}W=30$, not all high-energy  branches are shown.)}\label{plot:spectrumP_var}
 \end{center}
\end{figure}

\begin{figure}[hbpt]
 \begin{center}
  \includegraphics[width=\columnwidth]{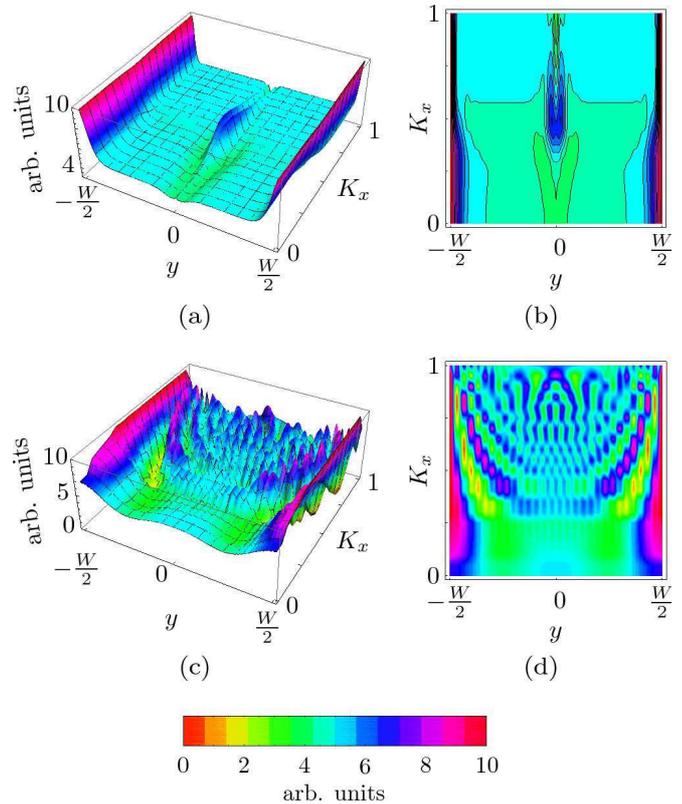}
   \caption{(Color online) Probability density of the Cooperon eigenmodes in the wire for $Q_\kt{SO}W/\pi=30$.
    (a) 3D plot, (b) density plot  for one of the two lowest branches,
    showing their edge mode character. (c) 3D plot and (d) density plot
    of  the  density of the third lowest mode, which shows bulk character.}\label{plot:CoopOrt2}
  \end{center}
\end{figure}

The exact diagonalization of the inverse  Cooperon propagator, as obtained
after the non-Abelian transformation, Eq.\,(\ref{transformed_Cooperon}),
 is performed in the basis
of transverse standing waves, satisfying Neumann boundary conditions, $\left\{1/\sqrt{W},\sqrt{2}/\sqrt{W}\cos\left((n\pi/W)(y-W/2)\right)\right\}$
with $n\in\mathbb{N}^*$, and  the plane waves  $\exp (i Q_x x)$ with momentum $Q_x$ along the wire.
The results of this calculation for different values of the dimensionless
wire width $Q_\kt{SO} W$ are shown in Fig.\,\ref{plot:spectrumP_var}. 
The numerical data points are attributed
to the different branches of the eigenenergy dispersion by comparing their eigenvectors. For small $Q_\kt{SO}W$,
the result is in accordance with the 0-mode approximation:
For small wire widths $W$, the z-component of the total spin,  $S_z$, is a good quantum number, 
as can be seen by expanding Eq.\,(\ref{A_s_transformed}) in $Q_\kt{SO} y$. Thus,  one can identify the  lowest modes with the transverse zero-modes of the  triplet modes corresponding to the eigenvalues of $S_z$, $m =0 , \pm 1$,
 in the rotated spin axis frame, denoting them as  $E_{\{t0,n=0\}}$ and $E_{\{t{\pm},n=0\}}$.
The minimum of the $E_{\{t0,n=0\}}$ mode is located at $Q_x=0$. The minimum of $E_{\{t-,0\}}$ is located at finite $Q_x>0$, in agreement with the 0-mode approximation. For larger $Q_\kt{SO}W$, the modes mix with respect to the spin quantum number and the   transverse quantization modes. As a consequence, energy level crossings which are present at small wire widths are lifted at larger widths, since the mixing of spin and transverse quantization   modes results in level repulsion being seen in Fig.\,\ref{plot:spectrumP_var} as avoided level crossings. The branches, $E_{\{t0,0\}}$ and $E_{\{t-,0\}}$, evolve into two modes which become degenerate at large values of $Q_\kt{SO}W$. These two modes are the only ones whose energy lies below the energy minimum which we obtained  for the 2D modes, $E/D_e=(7/16)Q_\kt{SO}^2$, for a finite $K_x$-interval around $K_x=0$. Therefore, we can identify these modes  with  edge modes which are created by the Neumann boundary conditions.  We can confirm that these are   edge modes  by considering their spatial distribution,  shown in Fig.\,\ref{plot:CoopOrt2}. 
Therefore, even in the limit of large widths $W$,  we do get in addition to the  spectrum obtained for  the 2D system with open boundary conditions case the edge modes, whose energy is lowered as seen in Fig.\,\ref{plot:spectrumP_var}. The presence of these edge states and the difference to the 2D system with open boundary conditions can be seen in  Appendix \ref{exactDiag} in the nondiagonal elements which are proportional to the width times the functions Eqs.\,(\ref{R-function1}) and (\ref{R-function2}). Even in the limit of wide wires there are nondiagonal matrix elements which give a significant contribution which cannot be neglected. The modes above $E/D_e=(7/16)Q_\kt{SO}^2$ are extended over the whole wire system and can thus be characterized as  bulk-states, as seen in Figs.\,\ref{plot:CoopOrt2} (c) and (d).\cite{Wenk:Diplomarbeit} 
\begin{figure}[hbtp]
    \begin{center}
        \includegraphics[width=\columnwidth]{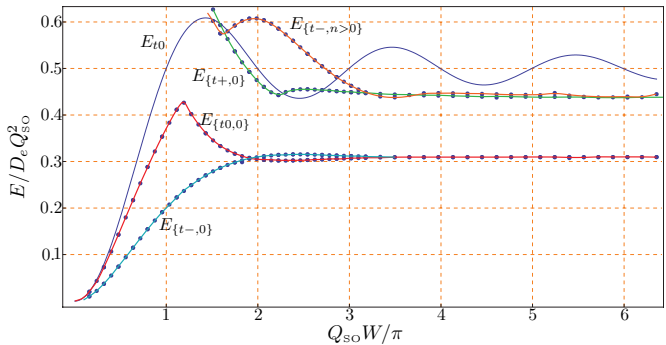}\\
        \caption{(Color online) Absolute minima of the lowest eigenmodes $E_{\{t0,0\}}$, $E_{\{t-,0\}}$, and
        $E_{\{t+,0\}}$ plotted as function of  $Q_\kt{SO}W/\pi= 2 W/L_\kt{SO}$.
         We note that  the minimum of $E_{\{t-,0\}} $ is  located  at $\pm K_x\neq 0$.  For comparison, the
        solution of the zero-mode approximation $E_{t0}$ is shown.}\label{plot:Wminima}
    \end{center}
\end{figure}
In Fig.\,\ref{plot:Wminima}, we compare the results which we obtained
in the 0-mode approximation  with the results of the   exact diagonalization. We plot the absolute minima of the spectra as function of the  dimensionless wire width parameter  $Q_\kt{SO}W/\pi = 2 W/L_\kt{SO}$. We confirm  the parabolic suppression of the lowest eigenvalues  for narrow wires  $\sim W^2/L_\kt{SO}^2$, obtained earlier.\cite{MalshukovPRB.61.R2413,Kettemann:PRL98:2007} We note that the oscillatory behavior of the triplet eigenvalues as function of W,  obtained in the 0-mode approximation,\cite{Kettemann:PRL98:2007} is diminished according to the  exact diagonalization. However, there remains a sharp  maximum of $E_{t0}$ at $Q_\kt{SO}W/\pi\approx 1.2$ and a shallow maximum of $E_{t-}$  at $Q_\kt{SO}W/\pi\approx 2.5$. As noted above, the values of the energy minima of $E_{t0}$ and $E_{t-}$ at larger widths $W$ are furthermore diminished as  a result of the edge mode character of these modes.
\subsubsection{Comparison to Solution of Spin-Diffusion Equation in Quantum Wires}
\begin{figure}[htbp]
     \begin{center}
         \includegraphics[width=3\columnwidth/4]{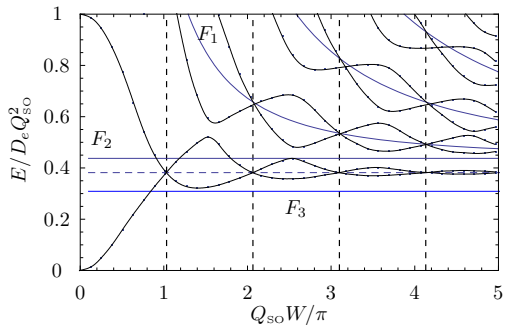}
         \caption{(Color online) Lowest eigenvalues at $K_x=0$ plotted against $Q_\kt{SO}W/\pi$. For comparison, the global minimum of the Cooperon spectrum for $Q_\kt{SO}W\gtrsim 9$ is plotted, $F_3$. Curves $F_1[n]$ are given by $7/16+\left((n/(Q_\kt{SO}W/\pi))\sqrt{15}/4\right)^2$, $n\in\mathbb{N}$. $F_2$ shows the energy minimum of the 2D case, $F_2\equiv F_1[n=0]$. Vertical dotted lines indicate the widths at which the lowest two branches degenerate at $K_x=0$. They are given by $n/(\sqrt{15}/4)$; consider that the wave vector for the minimum of the $E_{T_-}$ mode is $(\sqrt{15}/4)Q_\kt{SO}$.}\label{plot:NullstelleRaimondi}
     \end{center}
 \end{figure}

As shown above,  the  spin-diffusion operator and the triplet Cooperon propagator have the same
eigenvalue spectrum as soon as time-symmetry is not broken. Therefore, the minima of the spin-diffusion modes,
 which yield information on the spin relaxation rate, must be the same as the one of the triplet Cooperon propagator
as plotted in Fig.\,\ref{plot:Wminima}. In Ref. \onlinecite{Raimondi:2006}, the value at $K_x =0$, with $K_i=Q_i/Q_\kt{SO}$, has been plotted, as shown in Fig.\,\ref{plot:NullstelleRaimondi}. 
We note, however, that this does not correspond to the global minimum plotted in Fig.\,\ref{plot:Wminima}. The two lowest states exhibit two minima as can be seen in Fig.\,\ref{plot:spectrumP_var}: one local at $K_x=0$ and one global, which is for large  $Q_\kt{SO}W$  at $K_x\approx 0.88$. The first one is equal to the results given by Ref. \onlinecite{Raimondi:2006}. For the WL correction to the conductivity, however,  it is important to retain the global minimum, which is dominant in the integral over the longitudinal momenta.
\begin{figure}[t]
 \begin{center}
 \includegraphics[width=2\columnwidth/3]{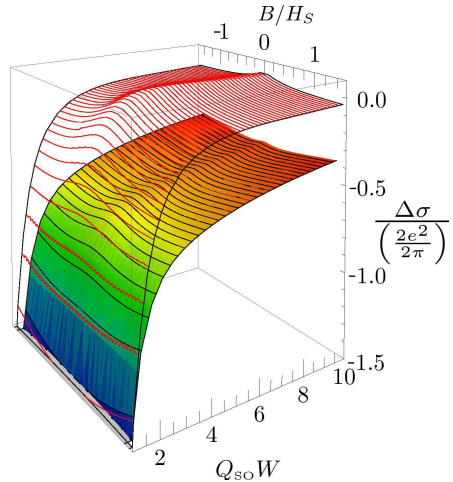}
\caption{(Color online) The quantum conductivity correction  in units of $2 e^2/2\pi$ as function of magnetic
 field $B$ (scaled with  bulk relaxation field $H_{s}$), and the wire width $W$ scaled with
  $1/Q_\kt{SO}$ for pure Rashba coupling and cutoffs $1/D_e Q_\kt{SO}^2\tau_\varphi=0.08$, $1/D_e Q_\kt{SO}^2\tau=4$: Comparison of the zero-mode calculation (grid without shading) to the exact diagonalization where  the lowest 21 triplet branches and seven singlet branches were taken into account.}\label{plot:DeltaSigma08Vergleich2}
\end{center}
\end{figure}
\subsubsection{Magnetoconductivity}
\begin{figure}[htbp]
\begin{center}
\includegraphics[width=3\columnwidth/4]{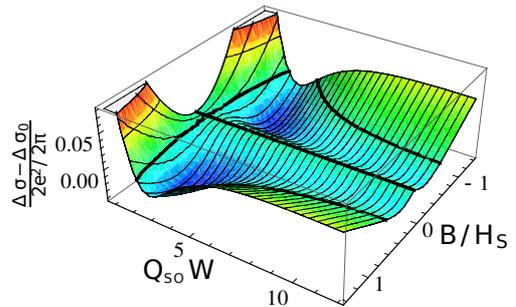}
\caption{(Color online) The relative magnetoconductivity $\Delta \sigma (B) - \Delta \sigma (B=0)$ in units of $2 e^2/2\pi$, with the same parameters and number of modes as in Fig.\,\ref{plot:DeltaSigma08Vergleich2}.}\label{plot:3conductanceC1gleich08Vergleich1}
\end{center}
\end{figure}

\begin{figure}[htbp]
\begin{center}
\includegraphics[width=3\columnwidth/4]{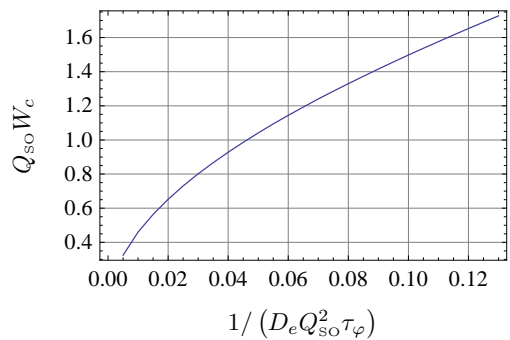}
        \caption{(Color online) Width of wire $W Q_\kt{SO}$
        at which there is a crossover from negative to positive
        magnetoconductivity
        as function of the lower cutoff
         $1/D_e Q_\kt{SO}^2 \tau_\varphi$.}\label{plot:sigmalowercutoff}
    \end{center}
\end{figure}

Now, we can proceed to calculate the quantum corrections to the conductivity using the exact diagonalization of the Cooperon propagator. In Fig.\,\ref{plot:DeltaSigma08Vergleich2}, we show the resulting conductivity as function of magnetic field and as function of the wire width $W$. Here, we have included for all wire widths the lowest seven singlet modes and the lowest 21 triplet modes. We choose this  number of modes so that  we included sufficient modes to describe correctly the widest wires considered with  $Q_\kt{SO}W = 10$. Thus,  for the considered low-energy cutoff, due to electron dephasing rate $1/\tau_{\varphi}$  of $1/D_e Q_\kt{SO}^2\tau_\varphi=0.08$ and the high energy cutoff $1/D_e Q_\kt{SO}^2\tau=4$ due to the elastic scattering rate,  we estimate that seven singlet modes fall in this energy range. Since for every transverse mode there are one singlet and three triplet modes, we therefore have to include 21 triplet modes, accordingly. We note a change from positive to negative magnetoconductivity as the wire width becomes smaller than the spin precession length $L_\kt{SO}$, in agreement with the results obtained within the 0-mode approximation, as reported earlier,\cite{Kettemann:PRL98:2007} plotted for comparison in Fig.\,\ref{plot:DeltaSigma08Vergleich2} (without shading). At the width, where the crossover occurs, there is a very weak magnetoconductance. This crossover width $W_c$ does depend on the lower cutoff, provided by the temperature-dependent dephasing rate $1/\tau_{\varphi}$. To estimate the dependence of $W_c$ on the dephasing rate, we have to analyze the contribution of each term in the denominator of singlet and triplet terms of the Cooperon. A significant change should arise if
\begin{equation}\label{crossoverEst}
\frac{1}{\tau_s}(W= W_c)=\frac{1}{\tau_\varphi}
\end{equation}
Assuming that this occurs for small wire widths, $Q_\kt{SO}W<1$, as confirmed for the
parameters we used, we apply Eq.\,(\ref{hso0}) to Eq.\,(\ref{crossoverEst}) and conclude that
\begin{equation}\label{Wc}
W_c\sim\frac{1}{\sqrt{\tau_\varphi}}
\end{equation}
If we calculate the crossover numerically in the 0-mode
approximation we get the relation plotted in Fig.\,\ref{plot:sigmalowercutoff} which coincides with Eq.\,(\ref{Wc}). 
We note that the change from weak antilocalization to weak localization may occur at a different width $W_c$ than 
 the  change of sign in the correction to the electrical conductivity $\Delta\sigma (B=0)$ occurs,
 $W_\kt{WL}$. 
However, we find that  the ratio $W_c/W_\kt{WL}$ is independent of the dephasing rate and the spin-orbit coupling strength $Q_\kt{SO}$.\\
Furthermore while there is quantitative  agreement with the 0-mode approximation in the magnitude of the magnetoconductivity
for all magnetic fields for small wire widths $W < L_\kt{SO}$, there is only qualitative
agreement at larger wire widths. In particular, the total magnitude of the conductivity
is reduced considerably in comparison with the 0-mode approximation.
We can attribute this to the reduction of the energy of the lowest Cooperon triplet modes
due to the emergence of edge modes, which is not taken into account
when neglecting transversal spatial variations, as is done in the 0-mode approximation.
Therefore, the 0-mode approximation overestimates the suppression of the
triple modes, resulting in an overestimate of the  conductivity. Similarly,
the magnetic field at which the magnetoconductivity changes its sign from
negative to positive is already at a  smaller magnetic field, as seen by the shift in the
minimum of the conductivity towards smaller magnetic fields (Fig.\,\ref{plot:3conductanceC1gleich08Vergleich1}), in comparison to the 0-mode approximation (unshaded)  in Fig.\,\ref{plot:DeltaSigma08Vergleich2}. This is in accordance with experimental observations, which showed clear deviations from the 0-mode approximation for larger wire widths, with a stronger magnetic field dependence than obtained in 0-mode approximation.\cite{hu05_1,hu05_2,hu05_3,hu05_4} Note that the nonmonotonous behavior of the triplet modes as function of the wire width, seen in  Fig.\,\ref{plot:spectrumP_var}, cannot be resolved in the width dependence of the conductivity.
\subsection{Other Types of Boundary Conditions}
\subsubsection{Adiabatic Boundary Conditions}
   When the lateral confinement potential V is smooth compared to the SO splitting, 
    that is, if $\lambda_F \partial_y V \ll \Delta_\kt{SO} = 2 k_F \alpha_2$, where $\lambda_F$ is the 
     Fermi wavelength, the  boundaries do not preserve  the spin, $s_\kt{in} \neq s_\kt{out}$, Eq.\,(\ref{bc}), since 
     the spin may adiabatically evolve as the electron is  scattered from such a smooth  boundary.\cite{PhysRevB.70.245310} If this applies,  the potential is adiabatic and  the spin  of the scattered electron 
     stays parallel to the field ${\bf B}_\kt{SO}$ as its momentum is changed.
 This leads to the boundary condition for the spin-density\cite{Raimondi:2006}
\begin{eqnarray}
s_x|_{y=\pm W/2}&=&0,\\
s_y|_{y=\pm W/2}&=&0,\\
\partial_y s_z|_{y=\pm W/2}&=&0.
\end{eqnarray}
We can transform this boundary condition to  the one of the triplet 
 Cooperon by using the unitary rotation  between the spin
density in the $s_i$ representation and the
triplet representation of the Cooperon,  $\tilde s_i$, Eq.\,(\ref{spinDensityTriplet}),
which leads to the boundary condition
\begin{eqnarray}
 \frac{1}{\sqrt{2}}(-s_x+\I s_y)|_{y=\pm W/2}=\tilde s_\upuparrows|_{y=\pm W/2}&=&0,\\
 \partial_ys_z|_{y=\pm W/2}=\partial_y\tilde s_\rightrightarrows|_{y=\pm W/2}&=&0,\\
 \frac{1}{\sqrt{2}}(s_x+\I s_y)|_{y=\pm W/2}=\tilde s_\downdownarrows|_{y=\pm W/2}&=&0.
\end{eqnarray} 
Now, if we require vanishing magnetization for the 1D case, then the diagonalization is done straightforwardly as already calculated in Ref. \onlinecite{Raimondi:2006}. We use a basis which satisfies the boundary conditions and therefore consists of $\sim$\!\,$\sin(qy)(1,0,0)^T$, $\sim$\!\,$\cos(qy)(0,1,0)^T$, and $\sim$\!\,$\sin(qy)(0,0,1)^T$, with $q=n\pi/W$, $n\in\mathbb{N}^*$. However, looking at the spin-diffusion operator [Eq.\,(\ref{Hsd})], we see immediately that if we set $\bf k$ to zero and use the fact that $s_{x,y}$ must vanish at the boundary and $s_{z}$ has to be constant for the chosen $\bf k$, we receive a polarized mode. Although this mode is a trivial solution, it differs from all other due to the fact that it has a finite spin relaxation time as $Q_\kt{SO}W$ vanishes. For the choice of basis for diagonalization, this means: 
We set $q_i=n_i\pi/W$ for respective $\tilde s_i$ therewith one state is described by $\{n_1,n_2,n_3\}=\{n,n+p_2,n+p_3\},\phantom{n} n\in \mathbb{N}^*,\phantom{n}p_2\in \{-1,0,1,\cdots\},\phantom{n} p_3\in \mathbb{N}$. In the case $\{n_1,n_2,n_3\}=\{n,n,n\}$ all branches diverge with reference to the eigenvalues in the limes of $Q_\kt{SO}W\rightarrow 0$, so that the spin relaxation time goes to zero for small wires.\cite{Raimondi:2006} In contrast, there is an additional branch in the case of $p_2=-1, p_3=0$ which has a finite eigenvalue and therefore finite spin relaxation time for small wire widths
\begin{equation}\label{adiabatic:E_finite}
\frac{E}{D_eQ_\kt{SO}^2}=2+K_x^2\left(1-\frac{32(Q_\kt{SO}W)^2}{\pi^4}\right)+\mathcal{O}[(Q_\kt{SO}W)^4].
\end{equation}
The smallest spin relaxation rate for vanishing $Q_\kt{SO}W$, $1/\tau_{s,1D}$, which is given for $k_x=0$, is found to be an eigenstate polarized in z direction which relaxes with the rate $1/\tau_{s,1D}=2/\tau_s$. It shows compared with the other modes a monotonous behavior as function of $Q_\kt{SO}W$.
If we allow magnetization for the 1D case, then the combination $p_1=-1,p_2=0$ leads to a valid solution. For wide wires the smallest absolute minimum is the 2D minimum $E/D_e=(7/16) Q_\kt{SO}^2$; there are no edge modes. But already at a width of $Q_\kt{SO}W=\pi/\sqrt{5}\approx 1.4$ all modes except the z-polarized exceed the rate $1/\tau_{s,1D}$.
\subsubsection{Tubular Wires}
 In tubular wires, such as carbon nanotubes, and InN nanowires in which only surface electrons conduct,\cite{Petersen2009} and radial core-shell InO nanowires,\cite{jung2008} the tubular topology of the electron system can be taken into account by periodic boundary conditions. In the following, we focus on wires where the dominant SO coupling is of Rashba type. If one requires furthermore that this SO-coupling strength is uniform and the wire curvature can be neglected,\cite{Petersen2009} the spectrum of the Cooperon propagator can be obtained by substituting in Eq.\,(\ref{eig_val_Coop2D_Tpm})  the transverse momentum $Q_y$ by the quantized values $Q_y = n 2\pi/W$, n is an integer, when $W$ is the circumference of the tubular wire. Thus, the spin relaxation rate remains unchanged, $1/\tau_s = (7/16)D_e Q_\kt{SO}^2 $. If then a magnetic field perpendicular to the cylinder axis is applied as done in Ref. \onlinecite{Petersen2009}, there remains a negative magnetoconductivity due to  the weak antilocalization, which  is enhanced due to the dimensional crossover from the 2D correction to the conductivity Eq.\,(\ref{2DG}) to the quasi-one-dimensional behavior of the quantum correction to the conductivity [Eq.\,(\ref{wl1D})]. In tubular wires in which the circumference fulfills the quasi-one-dimensional condition $W < L_{\varphi}$, the weak localization correction can then be written as
\begin{eqnarray}
&& \Delta \sigma =\nonumber\\
&&\frac{\sqrt{H_W}}{\sqrt{H_{\varphi}+ B^*(W)/4}} - \frac{\sqrt{H_W}}{\sqrt{ H_{\varphi}
 + B^*(W)/4 +  H_{s}(W) }}
 \nonumber \\
&&-2\frac{\sqrt{H_W}}{\sqrt{H_{\varphi}+ B^*(W)/4 + 7 H_{s}(W)/16}}
\end{eqnarray}
in units of $e^2/2 \pi$.
As in Eq.\,(\ref{wl1D}), we defined $H_W =1 /4 e W^2$, but the effective external magnetic field differs due to the different geometry:
Assuming that $W<l_B$, we have\cite{Kettemann:PRL98:2007}
\begin{align}
 \frac{1}{\tau_B}&=D_e(2e)^2B^2\langle y^2\rangle\\
&=D_e(2e)^2\frac{1}{2}\left(\frac{BW}{2\pi}\right)^2\\
\intertext{and the effective external magnetic field yields}
   B^*(W) &=(2e)\left(\frac{BW}{2\pi}\right)^2\label{beff_tubular}\\
&=(2e)(Br_\kt{tub})^2,
\end{align}
with the tube radius $r_\kt{tub}$.
The spin relaxation  field $H_{s}$ is  $H_s = 1/4 e D_e \tau_s$, with $1/\tau_{s} = 2 p_F^2  \alpha_2^2  \tau  $,
 or in terms of the effective Zeeman field $B_{\rm SO}$,
\begin{equation}
H_s = \frac{g \gamma}{16} \frac{B_{\rm so}(\epsilon_{\rm F})^2}{\epsilon_{\rm F}}.
\end{equation}
Thus the geometrical aspect, $\langle y^2\rangle_\kt{tube}/\langle y^2\rangle_\kt{planar}\approx 6.6$, might resolve the difference between measured and calculated SO coupling strength in Ref. \onlinecite{Petersen2009} where a planar geometry has been assumed to fit the data. This assumption leads in a tubular geometry to an underestimation of $H_s(W)$. The flux cancellation effect is as long as we are in the diffusive regime, $l_e\ll W$, negligible.
\section{Magnetoconductivity with Zeeman splitting}\label{Sec:MagnetoWithZeeman}
In the following, we want to study if  the Zeeman term, Eq.\,(\ref{zeemanTerm}), is modifying the magnetoconductivity. Accordingly, we assume that the magnetic field is perpendicular to the 2DES. Taking into account the Zeeman term to  first order in the external  magnetic field ${\bf B}=(0,0,B)^T$, the Cooperon is according to Eq.\,(\ref{Cooperon1}) given by
\begin{equation}
 \hat{C} ({\bf Q})= \frac{1}{D_e( {\bf Q} + 2 e {\bf A} + 2 e  {\bf A}_{\bf S})^2 + \I\frac{1}{2} \gamma (\bsigma^\prime-\bsigma)\mathbf{B}}.
\end{equation}
This is valid for magnetic fields  $\gamma B\ll 1/\tau$. 
Due to the term proportional to $(\bsigma^\prime-\bsigma)$, the singlet sector of the Cooperon mixes with the triplet one. We can find  the eigenstates of $C^{-1}$, $\ket{i}$ with the eigenvalues $1/\lambda_i$. Thus,  the sum over all spin up and down combinations $\alpha\beta,\beta\alpha$  in Eq.\,(\ref{DeltaSigmaQ}) for the 
 conductance correction simplifies in the singlet-triplet representation to
\begin{align}
 \sum_{\alpha\beta}C_{\alpha\beta\beta\alpha}={}&\sum_i\left(-\braket{\rightleftarrows}{i}\braket{i}{\rightleftarrows}+\braket{\upuparrows}{i}\braket{i}{\upuparrows}\right.\nonumber\\
&\left.+\braket{\rightrightarrows}{i}\braket{i}{\rightrightarrows}+\braket{\downdownarrows}{i}\braket{i}{\downdownarrows}\right)\lambda_i.
\end{align}
\subsection{2DEG}
\begin{figure}[htbp]
 \begin{center}
  \subfigure[]{\scalebox{1}{\includegraphics[width=3\columnwidth/4]{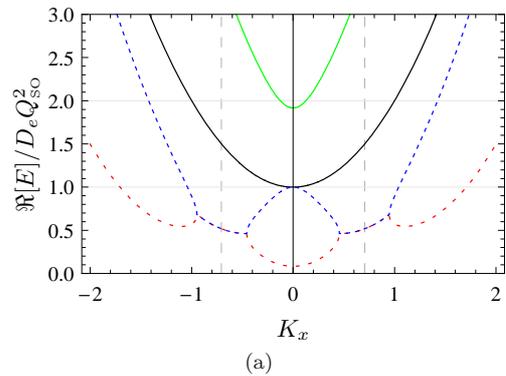}}}\\
\subfigure[]{\scalebox{1}{\includegraphics[width=3\columnwidth/4]{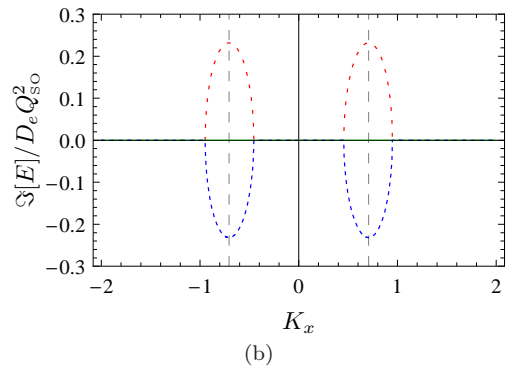}}}
   \caption{(Color online) (a) Real and (b) imaginary parts of the spectrum of the 2D Cooperon with Zeeman term of strength $g\mu_\kt{B}B/D_e Q_\kt{SO}^2=0.4$. $E_{\kt{B,2D},1}$ (black), $E_{\kt{B,2D},2}$ (red dashed), $E_{\kt{B,2D},3}$ (green), $E_{\kt{B,2D},4}$ (blue dashed). Dashed vertical lines are located at $K_x=\pm 1/\sqrt{2}$, the wave vector where the triplet mode $E_{T_-}$ and the singlet mode $E_S$ are crossing each other (without loss of generality $K_y=0$).}\label{plot:Zeeman31}
  \end{center}
\end{figure}

\begin{figure}[htbp]
 \begin{center}
\subfigure[]{\scalebox{1}{\includegraphics[width=2\columnwidth/3]{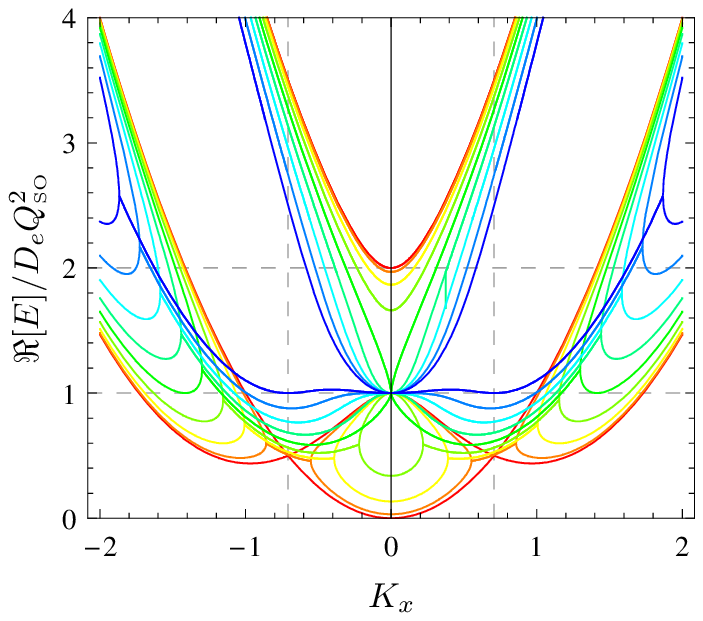}}}\\
\subfigure[]{\scalebox{1}{\includegraphics[width=2\columnwidth/3]{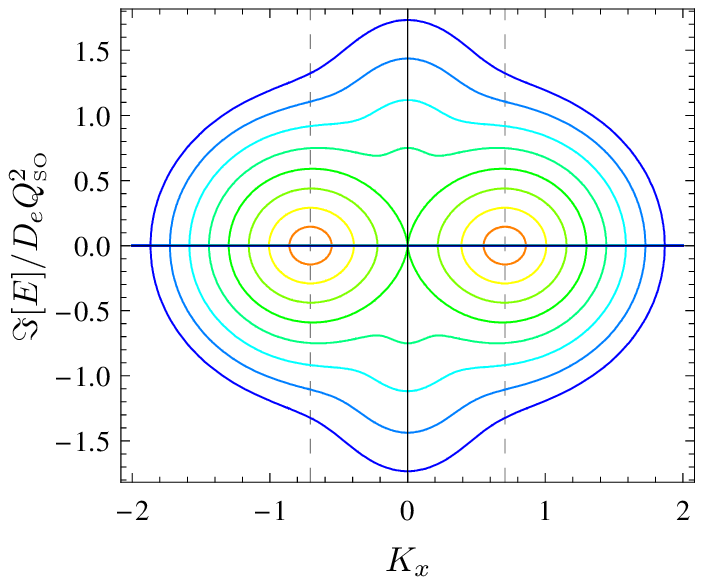}}}\\
\subfigure{\scalebox{.8}{\includegraphics[width=2\columnwidth/3]{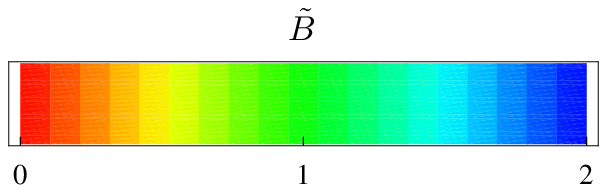}}}
   \caption{(Color online) (a) Real and (b) imaginary parts of the spectrum of the 2D Cooperon with Zeeman term of the strength $g\mu_\kt{B}B/D_e Q_\kt{SO}^2=0\dots 2$ in steps of $0.25$ (w.l.o.g $K_y=0$). The B independent mode $E_{T_0}$ is not shown.}\label{plot:Zeeman32}
  \end{center}
\end{figure}

The coupling of the singlet to the triplet sector  lifts the energy level crossings at $K=\pm 1/\sqrt{2}$ 
of the singlet $E_S$ and the triplet branch $E_{T_-}$ as can be seen in Fig.\,\ref{plot:Zeeman31} for a nonvanishing Zeeman coupling.
The spectrum, which is not positive definite anymore for all wave vectors, $K = Q/Q_{\rm SO}$, is given by
\begin{eqnarray}
 E_{\kt{B,2D},1\phantom{/4}}/D_eQ_\kt{SO}^2&=&1+K_x^2,\\
 E_{\kt{B,2D},2\phantom{/4}}/D_eQ_\kt{SO}^2&=&E_{\kt{B,2D},1}+\frac{f_1}{f_2^{1/3}}-\frac{1}{3}f_2^{1/3},\\
 E_{\kt{B,2D},3/4}/D_e Q_\kt{SO}^2&=&E_{\kt{B,2D},1}-\frac{1}{2}\frac{\left(1\pm \I\sqrt{3}\right) f_1}{f_2^{1/3}}\nonumber\\
&&\phantom{E_{\kt{B,2D},1}}+\frac{1}{6} \left(1\mp \I \sqrt{3}\right) f_2^{1/3},
\end{eqnarray}
with
\begin{eqnarray}
       f_1&=&\tilde{B}^2-4K_x^2-1,\\
       f_2&=&3\left(\sqrt{3} \sqrt{108 K_x^4+f_1^3}-18 K_x^2\right),\\
 \tilde{B}&=&g\mu_\kt{B}B/D_e Q_\kt{SO}^2,
\end{eqnarray}
Thus, there are spin states with the same real part of the Cooperon energy, so that they decay equally in time,  but the imaginary part is different, so that they precess with different frequencies around the magnetic field axis. A significant change of the Cooperon spectrum   appears when $g\mu_\kt{B}B/D_e$ exceeds $Q_\kt{SO}^2$, as can be seen in Fig.\,\ref{plot:Zeeman32}(b). All states with a low decay rate do precess now, due to a finite
imaginary value of their eigenvalue. Associated with this change is also a  change of the dispersion of the real part of $E_{\kt{B,2D},3}$ which changes for $K_y=0$ from a nearly quadratic dispersion in $K_x$, $a_0+a_1 K_x^2$ for $\tilde B<1$ to one which changes more slowly as $a_0+a_1 K_x^{2/3}+a_2 K_x^{4/3}+a_3K_x^2$ for $\tilde B\geq 1$ [see Fig.\,\ref{plot:Zeeman32} (a)].
\subsubsection*{Weak Field}
In the case of a weak Zeeman field, $\tilde B\ll 1$, the singlet and triplet sectors are still approximately  separated. 
 A finite $\tilde B\ll 1$ lifts however the energy of  the singlet mode to $E_{B,2D,2}(K=0)/D_eQ_\kt{SO}^2=\tilde{B}^2/2+\mathcal{O}(\tilde{B}^4)$, thus the singlet mode attains a finite gap, corresponding to a finite relaxation rate.
 The absolute minimum of two of  the triplet modes is also lifted by $E_{B,2D,2}(K=\pm\sqrt{15}/4)/D_eQ_\kt{SO}^2=7/16+(3/4)\tilde{B}^2+\mathcal{O}(\tilde{B}^4)$, while  their value is  independent of $\tilde B$ at $K=0$. In contrast, the minimum of the triplet mode $E_{B,2D,3}$, which approaches $E_{T_+}$ in the limit of no magnetic field (see Fig.\,\ref{plot:free_Hc}) is diminished to $E_{B,2D,3}(K=0)/D_eQ_\kt{SO}^2=2-\tilde{B}^2/2+\mathcal{O}(\tilde{B}^4)$. So, in summary, a weak Zeeman field renders all four Cooperon modes gapfull and that gap can  be interpreted as a finite relaxation rate or dephasing rate as the Zeeman coupling mixes all the spin states, breaking time-reversal invariance.
  \subsubsection*{Strong Field}
If we expand the spectrum in $1/\tilde{B}\ll 1$, we find that all modes have the same gap proportional to  the strength of the SO coupling, $D_eQ_\kt{SO}^2$, while two modes attain a finite imaginary part with opposite sign
\begin{eqnarray}
 E_{\kt{B,2D},1\phantom{/4}}/D_eQ_\kt{SO}^2&=&1+K_x^2,\\ 
E_{\kt{B,2D},2}/D_eQ_\kt{SO}^2&=&1+K_x^2+\mathcal{O}(1/\tilde B),\\
 E_{\kt{B,2D},3/4}/D_eQ_\kt{SO}^2&=&1+K_x^2\mp\I\tilde B+\mathcal{O}(1/\tilde B).
\end{eqnarray}
 Thus, a strong Zeeman field polarizes the spins and leads to their  precession. The SO interaction, which is too weak to flip the spins, merely results in a relaxation of all modes, corresponding to a dephasing of the spin precession.
\subsection{Quantum Wire with Spin-Conserving Boundary Conditions}
\begin{figure}[htbp]
 \begin{center}
\subfigure[]{\scalebox{1}{\includegraphics[width=2\columnwidth/3]{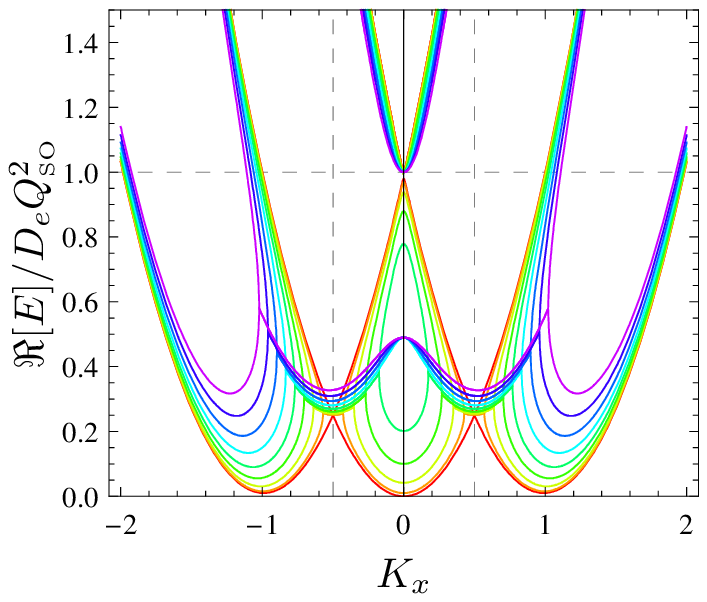}}}\\
\subfigure[]{\scalebox{1}{\includegraphics[width=2\columnwidth/3]{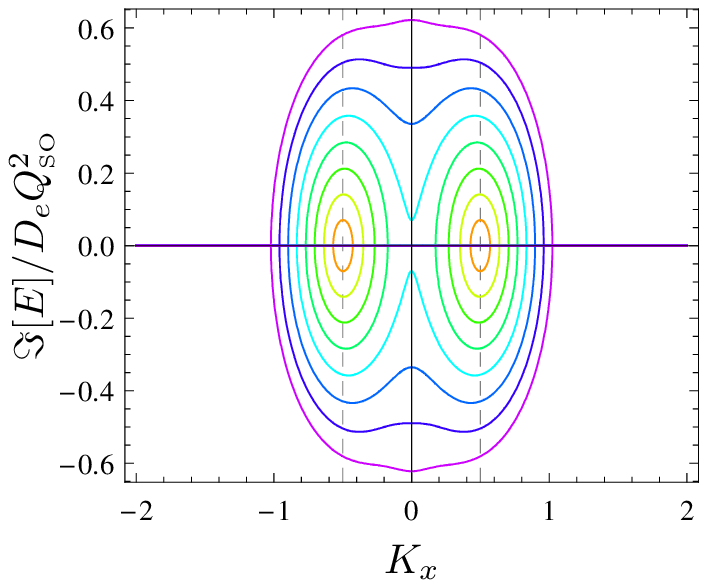}}}\\
\subfigure{\scalebox{0.8}{\includegraphics[width=2\columnwidth/3]{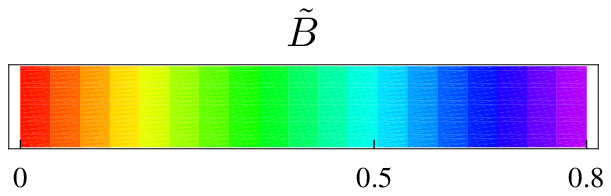}}}
   \caption{(Color online) (a) Real and (b) imaginary parts of the spectrum of the Cooperon with Zeeman term of the strength $g\mu_\kt{B}B/D_e Q_\kt{SO}^2=0\dots 0.8$ in steps of $0.1$ in a finite wire of the width $Q_\kt{SO}W=0.5$. The B independent mode $E_{t0}$ is not shown.}\label{plot:Zeeman42}
  \end{center}
\end{figure}

\begin{figure}[htbp]
 \begin{center}
   \includegraphics[width=3\columnwidth/4]{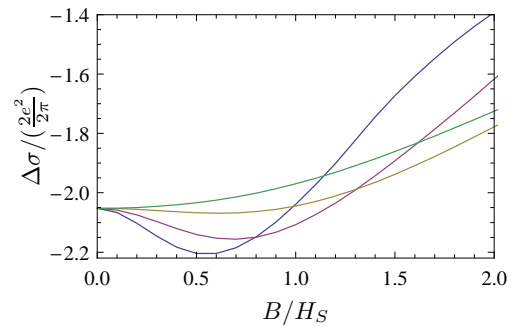}
\caption{(Color online) The magnetoconductance $\Delta\sigma(B)$ in a magnetic field perpendicular to the quantum well, where its coupling  via the Zeeman term is considered by exact diagonalization while its effect on the orbital motion is considered effectively by the  magnetic phase shifting rate $1/\tau_B(W)$ for wire width $Q_\kt{SO}W=1$, dephasing rate  $1/\tau_\varphi=0.06 D_e Q_\kt{SO}^2$, and elastic-scattering rate $1/\tau=4 D_e Q_\kt{SO}^2$. The strength of the contribution of the Zeeman term is varied by the material-dependent factor $\tilde{g}=g\mu_\kt{B} H_s/D_e Q_\kt{SO}^2$
in the range $\tilde{g}=0\dots 1.5$ in steps of $0.5$: The system changes from positive magnetoconductivity ($\tilde g=0$, green) to negative one ($\tilde g=0.5\dots 1.5$, continuous decrease of the absolute minimum).}\label{plot:SigmaZeeman}
  \end{center}
\end{figure}

\begin{figure}[htbp]
\begin{center}
\subfigure[]{\scalebox{1}{\includegraphics[width=3\columnwidth/4]{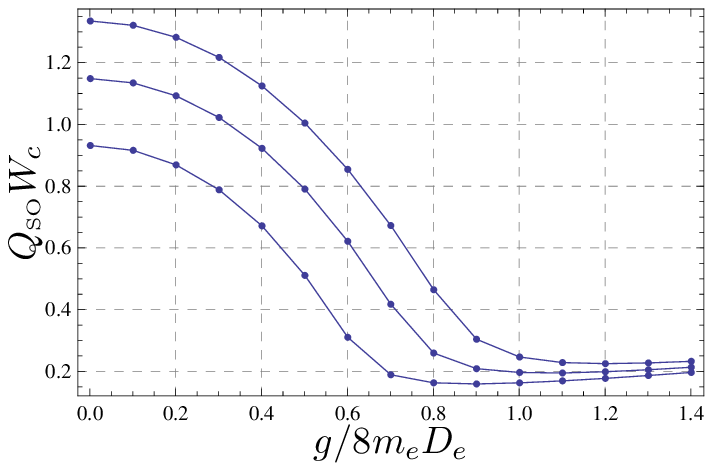}}}\\
\subfigure[]{\scalebox{1}{\includegraphics[width=3\columnwidth/4]{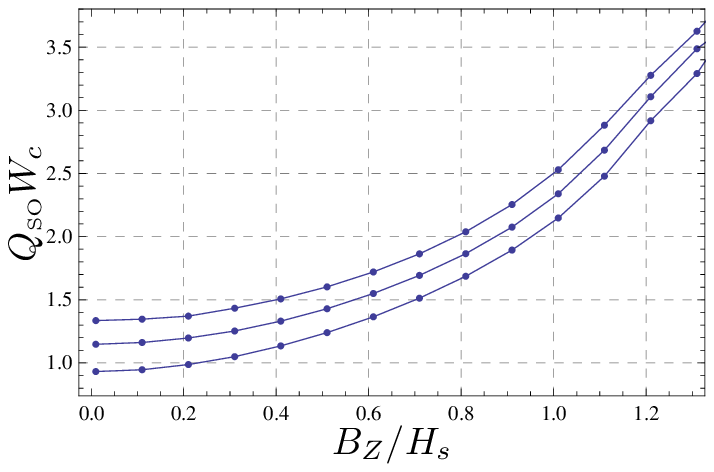}}}
  \caption{(Color online) (a) Change of crossover width $W_c$ with g factor: The magnetic field is included as an effective field $1/\tau_B$ and in the  Zeeman term. The strength of the contribution of the Zeeman term is varied by the material dependent factor $\tilde{g}=g\mu_\kt{B} H_s/D_e Q_\kt{SO}^2$.
(b) Change of crossover width $W_c$ with Zeeman field: 
To calculate $W_c$, we fix the Zeeman field to a certain value, horizontal axis, while we vary the effective field independently and calculate if negative or positive magnetoconductivity is present. For different Zeeman fields $B_Z/H_s$, we find thereby a different width $W_c$. Here, we set $g/8m_eD_e=1$. In (a) and (b), the cutoff due to dephasing is varied: $1/D_eQ_\kt{SO}^2\tau_\varphi=0.04,0.06,0.08$ (lowest first).
}\label{ShiftInMagnetoCrossover}
  \end{center}
\end{figure}

In the following, we want to study if a Zeeman field modifies the magnetoconductivity  and can shift the crossover from positive to negative Magnetoconductivity as function of wire width $W$. We have seen that for appropriate parameters the critical width $W_c$ is small compared with $L_\kt{SO}$. Therefore we stay in the 0-mode approximation to get a better overview of the physics. To do so, we first analyze the spectrum.\\
The modes with low decay rates are situated at $K_x=0$ and $K_x\approx\pm 1$ for small widths and small enough Zeeman field, $\tilde B\lesssim 1$, as can be seen in Fig.\,\ref{plot:Zeeman42}. For $K_x=0$, we have
\begin{align}
 E_{\kt{min},0}/D_e Q_\kt{SO}^2 ={}&\tilde{B}^2+\tilde{B}^4\left(1+\frac{(Q_\kt{SO}W)^2}{12}\right) \\
\intertext{and for $K_x=\pm 1$,}
 E_{\kt{min},\pm 1}/D_e Q_\kt{SO}^2 ={}& \frac{(Q_\kt{SO}W)^2}{24}+\tilde{B}^2\left(\frac{1}{2}-\frac{(Q_\kt{SO}W)^2}{96}\right)\nonumber\\
&+\tilde{B}^4\left(\frac{3}{16}-\frac{(Q_\kt{SO}W)^2}{384}\right).
\end{align}
As in the 2D case, we have a mode which is independent of the Zeeman field and the spectrum is equal to $E_{t0}$ with the eigenvector $(0,1,0,1)^T$. Using this spectrum, we estimate the correction to the static conductivity in the case of a magnetic field which we include by means of a Zeeman term together with an effective magnetic field appearing in the cutoff $1/\tau_B$ as described in Sec. \ref{Sec:QuantTranspCor}. The $\tilde{g}=g/8m_eD_e$ factor is used as a material-dependent parameter. In Fig.\, \ref{plot:SigmaZeeman}, we see that for large enough $\tilde g$ factor, the system changes from positive
magnetoconductivity{\textemdash}in the case without Zeeman field and a small-enough wire width{\textemdash}to negative magnetoconductivity at a finite Zeeman field for the same wire. Hence, the ratio $W_c/W_\kt{WL}$ changes and one has to be careful not to confuse the crossover defined by a change of the sign of the quantum correction, WL$\rightarrow$WAL, and the crossover in the magnetoconductivity. To give an idea how the crossover $W_c$ depends on $\tilde g$ and the strength of the Zeeman field we analyze two different systems as plotted in Fig.\,\ref{ShiftInMagnetoCrossover}: The first one, plot (a), shows the drop of $W_c$ in a system as just described where we have one magnetic field which we include with an orbital and a Zeeman part. For small $\tilde g$ we have $Q_\kt{SO}W_c(\tilde g)=Q_\kt{SO}W_c(\tilde g=0)-\text{const\,}{\tilde g}^2$, where const is about 1 in the considered parameter space. In the second system [Fig.\,\ref{ShiftInMagnetoCrossover}(b)], we assume that we can change the orbital and the Zeeman field separately. The critical width is plotted against the Zeeman field. To calculate $W_c$, we fix the Zeeman field to a certain value, horizontal axis in plot (b), while we vary the effective field and calculate if negative or positive magnetoconductivity is present. For different Zeeman fields $B_Z/H_s$ we get different $W_c$. We see that $W_c$ is shifted to larger widths as the Zeeman field is increased, $Q_\kt{SO}W_c(B_Z/H_s)=Q_\kt{SO}W_c(B_Z=0)+\text{const\,}(B_Z/H_s)^2$, where const is about 1 in the considered parameter space, while $\Delta\sigma(1/\tau_\kt{B}=0)$ (not plotted) is lowered as long as we assume small Zeeman fields. If we notice that $B_Z$ mixes singlet and triplet states it is understood that there is no gapless singlet mode anymore and therefore $\Delta\sigma(1/\tau_\kt{B}=0)$ must decrease for low Zeeman fields.\\
To estimate $\tilde g$, we  take typical values for a GaAs/AlGaAs system 
and assume the electron density to be $n_s=1.11\times 10^{11} \text{\,cm}^{-2}$, the effective mass $m_e/m_{e_0}=0.063$, the Land\'{e} factor $g=0.75$ and an elastic mean-free path of $l_e=10$ nm in a wire with $Q_\kt{SO}W=1$,  corresponding to $W=1.2 \text{\,}\mu$m,  if we assume a Rashba spin-orbit coupling strength of $\alpha_2=5$ meV\AA. We thus get $\tilde g\approx 0.1$ and find that the  Zeeman coupling due to the perpendicular magnetic field can have a measurable, albeit small effect on the magnetoconductance  in GaAs/AlGaAs systems.
\section{Conclusions}\label{Sec:conclusions}
In conclusion, in wires with spin-conserving boundaries and a width $W$  smaller than bulk spin precession length
$L_\kt{SO}$, the spin relaxation  due to linear Rashba SO coupling is suppressed according to the spin relaxation rate 
$(1/ \tau_s) (W) =(\pi^2/3)\left(W/L_\kt{SO}\right)^2 (1/ \tau_s)$, where $1/\tau_{s} = 2 p_F^2  \alpha_2^2 \tau  $.
The enhancement of  spin relaxation length $L_s = \sqrt{D_e \tau_s (W)}$ can be understood as follows: The area an electron covers by diffusion in  time $\tau_s$ is  $W L_s$.
This spin relaxation occurs if that area is equal $L_\kt{SO}^2$,\cite{falko} which  yields $1/L_s^2 \sim 1/D_e \tau_s\sim(W/L_\kt{SO})^2/L_\kt{SO}^2$, in agreement with Eq.\,(\ref{hso0}).
For larger wire widths, the exact diagonalization reveals a nonmonotonic behavior of the spin relaxation as function of the wire width of the long-living eigenstates. The spin relaxation rate is first enhanced before it is suppressed as the widths $W$ is decreased. The longest living modes are found to exist at the boundary of wide wires. Since we identified  a direct transformation from the spin-diffusion equation to the Cooperon equation,  we could show that these edge modes affect the conductivity:  the 0-mode approximation overestimates the conductivity for larger wire widths $Q_\kt{SO}W>1$ since it does not take into account these edge modes. They add a larger  contribution to the negative triplet term of the quantum correction than the bulk modes do, since they relax more slowly. This also results in  a shift in the minimum of the conductivity towards smaller magnetic fields in comparison to the 0-mode approximation. The reduction of  spin relaxation has recently been observed in optical measurements of \textit{n}-doped  InGaAs quantum wires\cite{holleitner}  and in transport measurements.\cite{hu05_1,hu05_2,hu05_3,hu05_4,gh05} Recently in Ref. \onlinecite{kunihashi:226601}, the enhancement of spin lifetime due to dimensional confinement in gated InGaAs wires with gate controlled SO coupling was reported.
Reference \onlinecite{holleitner} reports saturation of  spin relaxation  in narrow wires, $W \ll L_\kt{SO}$, attributed to  cubic Dresselhaus coupling.\cite{Kettemann:PRL98:2007} The contribution  of  the linear and cubic Dresselhaus SO interaction to the spin relaxation turns out to depend strongly on growth direction and will be studied in more detail in Ref. \onlinecite{wenk2}.
 Including both the linear Rashba and Dresselhaus SO coupling we have  shown that there exist two long persisting spin helix solutions in narrow wires even for arbitrary strength of both SO coupling effects. This is in contrast to the 2D case, where the condition $\alpha_1=\alpha_2$, respectively in the case where the cubic Dresselhaus term cannot be neglected, $\alpha_1-m_e\gamma\epsilon_F/2=\alpha_2$, is required to find persistent spin helices \cite{bernevig,liu:235322} as it was measured recently (Ref. \onlinecite{Koralek2009}).
Regarding the type of boundary, we found that the injection of polarized spins into nonmagnetic material is favorable for wires with a smooth confinement, $\lambda_F \partial_y V < \Delta_\kt{SO} = 2 k_F \alpha_2$. With such adiabatic boundary conditions,  states which are polarized in z direction relax with a finite rate for wires with widths $Q_\kt{SO}W\ll 1$, while the spin relaxation rate of all other  states diverges in that limit. 
 In tubular wires with periodic boundary conditions, the spin relaxation is found to remain constant as the wire circumference is reduced. Finally, by including the Zeeman coupling to the perpendicular magnetic field,  we have shown that for spin-conserving boundary condition the critical wire width, $W_c$, where the crossover from negative to positive magnetoconductivity occurs, depends not only on the dephasing rate but also depends on the g factor of the material.
\begin{acknowledgments}
We thank M. Kossow and M. Milletari for stimulating discussions. P.W. thanks the Asia Pacific Center for Theoretical Physics for hospitality. This research was supported by DFG-SFB Project No. 508 B9 and by WCU (World Class University) program through the Korea Science and Engineering Foundation funded by the Ministry of Education, Science and Technology (Project No. R31-2008-000-10059-0).
\end{acknowledgments}
\appendix
\section{Spin-Conserving Boundary}\label{derivationBoundary}
In the following we set $\gamma=1$. In order to generate a finite system, we need to specify the  boundary conditions. These can be different for the  spin and charge current. Here we derive the spin-conserving boundary conditions. Let us first recall the classical derivation of the  diffusion current  density in the  wire at the position ${\bf r}$. The current  density ${\bf j} $  at position ${\bf r}$ can be related by continuity  to  all currents in its vicinity which are directed towards that position. Thus, ${\bf j} ({\bf r},t) = \langle {\bf v} \rho ({\bf r} - \Delta {\bf x}) \rangle$ where an angular average over all possible directions of the velocity ${\bf v}$ is taken. Expanding in $\Delta {\bf x} = l_e {\bf v}/v$ and noting that $\langle {\bf v}  \rho ({\bf r}) \rangle =0$, one gets ${\bf j} ({\bf r},t) = \langle {\bf v} (- \Delta {\bf x} ) \mathbf{\nabla} \rho ({\bf r}) \rangle = -(v_F l_e/2)  \mathbf{\nabla} \rho ({\bf r})  = - D_e \mathbf{\nabla} \rho ({\bf r}) $. For the classical spin-diffusion current of  spin component $s_i$, as defined by ${\bf j}_{s_i} ({\bf r},t) = {\bf v} s_i ({\bf r} ,t)$, there is the complication that the spin keeps precessing as it moves from ${\bf r} - \Delta {\bf x}$ to $ {\bf r} $, and that the SO field changes its direction with the direction of the electron velocity ${\bf v}$. Therefore, the \textit{0}th order term in the expansion in $\Delta {\bf x} $ does not vanish, rather, we get
\begin{equation}\label{spincurrent1}
 {\bf j}_{s_i} ({\bf r},t) = \langle {\bf v} s_i^{\bf k} ({\bf r} ,t) \rangle - D_e\mathbf{\nabla} s_i ({\bf r},t),
\end{equation}
where  $s_i^{\bf k} $ is the part of the spin-density which evolved from the spin-density at ${\bf r} - \Delta {\bf x}$ moving with velocity ${\bf v}$ and momentum ${\bf k}$. Noting that the spin precession on ballistic scales $t \le \tau$ is governed by the Bloch equation
\begin{equation} \label{ballistic}
\frac{\partial {\bf \hat{s}}}{\partial t} = {\bf \hat{s}} \times    {\bf  B_\kt{SO} ({\bf k}) } 
- \frac{1} {\hat{\tau}_s} {\bf \hat{s}},
\end{equation}
we find by integration of Eq.\,(\ref{ballistic}) that after the scattering time $\tau$, 
 the spin-density components  are given by $s_i^{\bf k} =-\tau \left[ {\bf B}_\kt{SO} ({\bf k})  \times {\bf s} \right]_i $ so that we can rewrite the first term in Eq.\,(\ref{spincurrent1}) yielding the total spin-diffusion current as
\begin{equation} \label{spincurrent}
{\bf j}_{s_i} = - \tau \langle  {\bf v}_F \left[{\bf B}_\kt{SO} ({\bf k})  \times {\bf s} \right]_i \rangle
-D_e\mathbf{\nabla} s_i.
\end{equation}
In this section, we consider specular scattering from the boundary with  the  condition that  the 
 spin is conserved, so that the spin current density normal to the boundary must vanish
\begin{equation}
 {\bf n}\cdot{\bf j}_{s_i}|_{\pm W/2}=0,\label{boundaryDiffusion}
\end{equation}
where ${\bf n}$ is the vector normal to the boundary. Noting the relation between the spin-diffusion equation in the $s_i$ representation and the triplet components of the Cooperon density  $\tilde s_i$ ($\{\ket{\upuparrows},\ket{\rightrightarrows},\ket{\downdownarrows}\}$), Eq.\,(\ref{UHSD}),
\begin{align}
& U_\kt{CD}(\epsilon_{ijk}B_{\kt{SO},j})_{i=1..3,k=1..3}U_\kt{CD}^\dagger\nonumber\\
&=-\I(\bra{\tilde s_i}{\bf B}_\kt{SO}\cdot{\bf S}\ket{\tilde s_k})_{i=1..3,k=1..3},
\end{align}
where the matrix $U_\kt{CD}$ is given by Eq.\,(\ref{UHSDrepresent}), we can thereby transform  the boundary condition for the spin-diffusion current, Eq.\,(\ref{boundaryDiffusion}), to the triplet components of the Cooperon density  $\tilde s_i$,
\begin{align}
0= {\bf n}\cdot{\bf j}_{\tilde s_i}|_{y= \pm W/2}.
\end{align}
 Requiring also that the charge density is vanishing normal to the transverse boundaries, 
 which transforms into the condition $-\I\partial_{\bf n} \tilde \rho|_\kt{Surface} =0 $
 for the singlet component of the Cooperon density $\tilde \rho$,
 we
 finally  get the boundary conditions for the Cooperon without external magnetic field,  Eq.\,(\ref{bc}),
\begin{align}
\left(-\frac{\tau}{D_e}{\bf n}\cdot\langle{\bf v}_F [{\bf B}_\kt{SO}({\bf k})\cdot{\bf S}]\rangle-\I\partial_{\bf n}\right) C|_\kt{Surface} &=0.
\intertext{The last expression can be rewritten using the effective vector potential ${\bf A_S}$, Eq.\,(\ref{Cooperon1}),}
\left({\bf n}\cdot2e{\bf A_S}-\I\partial_{\bf n}\right) C|_\kt{Surface} &=0.
\end{align}
In the case of Rashba and linear and cubic Dresselhaus SO coupling in $(001)$ systems, we get
\begin{eqnarray}\label{AsBso}
 &&\frac{D_e}{\tau}2e{\bf A_S}=-\langle {\bf v}_F ({\bf B}_\kt{SO}({\bf k})\cdot{\bf S})\rangle\nonumber\\
&&=v_F^2 m_e \left(
\begin{array}{cc}
 -(\alpha_1-\frac{\gamma_D(m_ev_F)^2}{4}) & -\alpha_2\\
\alpha_2 & \alpha_1-\frac{\gamma_D(m_ev_F)^2}{4}
\end{array}
\right).{\bf S}.\nonumber\\
&&
\end{eqnarray}
\section{Relaxation Tensor}\label{RelaxTensor}
To connect the effective vector potential ${\bf A_S}$ with the spin relaxation tensor, we notice that $\hat{\tau}$ can be rewritten in the following way:
\begin{align}
\frac{1}{\hat{\tau}_s}={}& \tau (\langle {\bf B}_\kt{SO}({\bf k})^2\rangle
\delta_{ij}-\langle B_\kt{SO}({\bf k})_i B_\kt{SO}({\bf
k})_j\rangle )_{i=1..3,j=1..3}\\
\intertext{using $U_\kt{CD}$, Eq.\,(\ref{UHSD}),}
={}&\tau U_\kt{CD}^\dagger\{\langle B_\kt{SO}({\bf k})_x\rangle
S_x^2+\langle B_\kt{SO}({\bf k})_y\rangle S_y^2\nonumber\\
&+\langle
B_\kt{SO}({\bf k})_x
B_\kt{SO}({\bf k})_y\rangle (S_xS_y+S_yS_x)\}U_\kt{CD}\\
={}&\tau U_\kt{CD}^\dagger\langle ({\bf B}_\kt{SO}({\bf
k}).{\bf S})^2 \rangle U_\kt{CD}\\
={}&\frac{\tau}{v_F^2} U_\kt{CD}^\dagger\langle ({\bf v}_F{\bf
B}_\kt{SO}({\bf k}).{\bf S})^2 \rangle U_\kt{CD}.
\end{align}
Because
\begin{equation}
 \frac{\tau}{v_F^2}\langle ({\bf v}_F [{\bf B}_\kt{SO}({\bf k}).{\bf S}])^2\rangle=\frac{2\tau}{v_F^2}\langle ({\bf v}_F [{\bf B}_\kt{SO}({\bf k}).{\bf S}])\rangle^2
\end{equation}
is true for linear Rashba and linear Dresselhaus SO coupling but, in general, false if cubic-in-k terms are included in the SO field, we have to write
\begin{equation}
\tau\langle ({\bf
B}_\kt{SO}({\bf k}).{\bf S})^2\rangle = \frac{2\tau}{v_F^2}\langle ({\bf v}_F{\bf
B}_{\kt{SO}}({\bf k}).{\bf S}) \rangle^2+\text{ct}
\end{equation}
so that we conclude
\begin{equation}\label{spinRelaxTensorVektorfeld}
\frac{1}{\hat{\tau}_s}=U_\kt{CD}^\dagger(D_e(2e {\bf A}_{{\bf
S}})^2+ct)U_\kt{CD}
\end{equation}
with the separated cubic part $ct=D_e
m_e^2\epsilon_F^2\gamma_D^2(S_x^2+S_y^2)$. This reflects
nothing but the fact that the effective SO Zeeman term in Eq.\,(\ref{hamiltonian}) can only be rewritten as a vector potential ${\bf A_S}$ when the SO coupling is linear in momentum.
\section{Weak Localization Correction in 2D}\label{weakLoc2D}
\begin{figure}[htbp]
\begin{center}
\includegraphics[width=3\columnwidth/4]{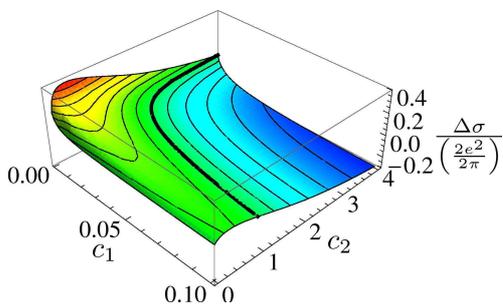}
   \caption{(Color online) Weak localization correction in 2D in units of $(2e^2/2\pi)$. The parameters are $c_1=1/D_e Q_\kt{SO}^2\tau_\varphi$
        and $c_2=1/D_e Q_\kt{SO}^2\tau$. Thick line indicates $\Delta\sigma=0$.}\label{correction2d}
    \end{center}
\end{figure}

In contrast to the case where we have a wire with a finite width, we can calculate the weak localization correction to the conductivity analytically in the 2D case.
The cutoffs due to dephasing $c_1=1/D_e
Q_\kt{SO}^2\tau_\varphi$ and elastic scattering $c_2=1/D_e
Q_\kt{SO}^2\tau$ determine whether we have a positive or negative correction. Integrating over all possible wave vectors
$K=k/Q_\kt{SO}$ in the case without boundaries yields
\begin{widetext}
\begin{eqnarray}
\Delta\sigma &=&-\frac{2 e^2}{2\pi}\frac{1}{(2\pi)^2}
\int_0^{\sqrt{c_2}}dK(2\pi K)\left(-\frac{1}{E_S(Q_\kt{SO}K)/Q_\kt{SO}^2+c_1}+\frac{1}{E_{T_0}(Q_\kt{SO}K)/Q_\kt{SO}^2+c_1}\right.\nonumber\\
&& \left.+\frac{1}{E_{T_+}(Q_\kt{SO}K)/Q_\kt{SO}^2+c_1}+\frac{1}{E_{T_-}(Q_\kt{SO}K)/Q_\kt{SO}^2+c_1}\right)
\end{eqnarray}
\begin{eqnarray}
\phantom{\Delta\sigma}&=&-\frac{2 e^2}{2\pi}\left(-\frac{1}{2}\ln\left(1+\frac{c_2}{c_1}\right)+\frac{1}{2}\ln\left(1+\frac{c_2}{1+c_1}\right)\right.\nonumber\\
&&+\left\{\frac{\arctan\left(\frac{5}{4}\frac{1}{\sqrt{\frac{7}{16}+c_1}}\right)-\arctan\left(\frac{\sqrt{\frac{1}{16}+c_2}+1}{\sqrt{\frac{7}{16}+c_1}}\right)}{\sqrt{\frac{7}{16}+c_1}}-\frac{1}{2}\ln\left(\frac{2+c_1}{\frac{3}{2}+c_1+c_2+2\sqrt{\frac{1}{16}+c_2}}\right)\right\}\nonumber\\
&&+\left\{\frac{\arctan\left(\frac{3}{4}\frac{1}{\sqrt{\frac{7}{16}+c_1}}\right)+\arctan\left(\frac{\sqrt{\frac{1}{16}+c_2}-1}{\sqrt{\frac{7}{16}+c_1}}\right)}{\sqrt{\frac{7}{16}+c_1}}-\frac{1}{2}\ln\left(\frac{1+c_1}{\frac{3}{2}+c_1+c_2-2\sqrt{\frac{1}{16}+c_2}}\right)\right\}\nonumber\\
&&\left.\right). \label{2DG}
\end{eqnarray}
\end{widetext}
As an example, we choose parameters which have been used in the case
of boundaries, $1/D_e Q_\kt{SO}^2\tau_\varphi=0.08, 1/D_e
Q_\kt{SO}^2\tau=4$:  $\Delta\sigma/(2e^2/2\pi)=-0.29$. The exact calculation of wide wires ($Q_\kt{SO}W>1$) approaches this limit as can be seen in Fig.\,\ref{plot:DeltaSigma08Vergleich2}. 
The weak localization correction in 2D as function of these
parameters is plotted in Fig.\,\ref{correction2d}.
\section{Exact Diagonalization}\label{exactDiag}
We  write the inverse Cooperon propagator, the Hamiltonian $\tilde{H}_{c}$, in the representation of the longitudinal momentum $Q_x$, the  quantized transverse momentum with quantum number $n\in\mathbb{N}$,  and in the representation of singlet and  triplet states with quantum numbers $S,m$, where we note that  $\tilde{H}_{c}$ is diagonal in $Q_x$,
\begin{equation}
\langle Q_x, n, S, m  \mid  \tilde{H}_{c} \mid Q_x, n', S', m' \rangle.
\end{equation}
The spin subspace is  thus represented by  $4\times 4$ matrices, which we order starting with the singlet $S=0$ and then $S=1,m=1$, $m=0$, and $m=-1$. Thus, we get
\begin{equation}
         \langle Q_x, n \mid  \tilde{H}_{c} \mid Q_x, n \rangle =  Q_\kt{SO}^2 \left(
       \begin{array}{cccc}
         A_n & 0 & 0 & 0   \\
         0 & B_n & \I F_n & D_n \\
         0 & -\I F_n & C_n & \I F_n \\
         0 & D_n & -\I F_n & B_n     \\
       \end{array}
     \right).
\end{equation}\label{UCooperonDiagonalMatrix}
The calculation of the matrix elements yields (we set $P = Q_\kt{SO} W/\pi$)
\begin{eqnarray}
    A_0&=&K_x^2,\\
    B_0&=&\frac{3}{4}+K_x^2-\frac{1}{4}\frac{\sin(P \pi)}{P\pi},\\
    C_0&=&\frac{1}{2}+K_x^2+\frac{1}{2}\frac{\sin(P \pi)}{P\pi},\\
    D_0&=&-\frac{1}{4}\enspace\enspace\enspace-\frac{1}{4}\frac{\sin(P \pi)}{P\pi},\\
    F_0&=&\enspace\enspace\enspace\enspace\enspace\sqrt{2}K_x\frac{\sin(\frac{P\pi}{2})}{\frac{P\pi}{2}},
\end{eqnarray}
and for $n>0$:
\begin{eqnarray}
    A_n&=&\phantom{\frac{3}{4}+} K_x^2+\left(\frac{n}{P}\right)^2,\\
    B_n&=&\frac{3}{4}+K_x^2+\left(\frac{n}{P}\right)^2+\frac{2P^2-n^2}{4(n+P)(n-P)}\frac{\sin(P\pi)}{P\pi},\nonumber\\
&&
\end{eqnarray}
\begin{eqnarray}
    C_n&=&\frac{1}{2}+K_x^2+\left(\frac{n}{P}\right)^2-\frac{2P^2-n^2}{2(n+P)(n-P)}\frac{\sin(P\pi)}{P\pi},\nonumber\\
&&\\
    D_n&=&-\frac{1}{4}-\frac{n^2-2P^2}{4(n-P)(n+P)}\frac{\sin(P\pi)}{P\pi},\\
    F_n&=&\frac{\sqrt{2}(2n^2-P^2)}{2\left(n-\frac{P}{2}\right)\left(n+\frac{P}{2}\right)}\frac{\sin(\frac{P\pi}{2})}{\frac{P\pi}{2}}.
\end{eqnarray}
For $n\neq n'$, the spin matrices have the form
\begin{equation}
      \langle Q_x, n \mid  \tilde{H}_{c} \mid Q_x, n' \rangle  = \frac{Q_\kt{SO}^2}{\pi}P  \left(
         \begin{array}{rrrr}
            0  & 0 & 0 & 0\\
            0& a & \I g & d \\
            0& -\I g & b & \I f \\
            0& d & -\I f & c \\
         \end{array}
       \right).
\end{equation}
Calculating the matrix elements  for $n=0,n'>0$, we get
\begin{widetext}
\begin{eqnarray}
 a &=& \frac{1}{\sqrt{2}}\left(\frac{\left(1+(-1)^{n'}\right) \sin(P \pi
   )}{(n'-2 P) (n'+2 P)}-\frac{4
   \left(-1+(-1)^{n'}\right) K_x \cos
   \left(\frac{P \pi }{2}\right)}{(n'-P)
   (n'+P)}\right), \\
b &=&  -\frac{\sqrt{2} \left(1+(-1)^{n'}\right) \sin (P
   \pi )}{(n'-2 P) (n'+2 P)},\\
c &=& \frac{1}{\sqrt{2}}\left(\frac{4 \left(-1+(-1)^{n'}\right) K_x
   \cos \left(\frac{P \pi }{2}\right)}{(n'-P)
   (n'+P)}+\frac{\left(1+(-1)^{n'}\right)
   \sin (P \pi )}{(n'-2 P) (n'+2
   P)}\right),
\end{eqnarray}
\begin{eqnarray}
d &=& \frac{\left(1+(-1)^{n'}\right) \sin (P \pi
   )}{\sqrt{2} (n'-2 P) (n'+2 P)}, \\
f &=&\phantom{-} 2 \left(\frac{\left(-1+(-1)^{n'}\right) \cos (P
   \pi )}{2 (n'-2 P) (n'+2
   P)}-\frac{\left(1+(-1)^{n'}\right) K_x
   \sin \left(\frac{P \pi }{2}\right)}{(n'-P)
   (n'+P)}\right),\\
  g &=& -2 \left(\frac{\left(-1+(-1)^{n'}\right) \cos (P
   \pi )}{2 (n'-2 P) (n'+2
   P)}+\frac{\left(1+(-1)^{n'}\right) K_x
   \sin \left(\frac{P \pi }{2}\right)}{(n'-P)
   (n'+P)}\right).
\end{eqnarray}
And  for $n>0,n'>0$, we get
\begin{eqnarray}
  a &=& R_{\{2,+\}} \sin (P \pi )+4 K_x
   R_{\{1,-\}} \cos \left(\frac{P \pi
   }{2}\right),\\
 b &=&-2 R_{\{2,+\}}
   \sin (P \pi ), \\
  c &=&R_{\{2,+\}} \sin
   (P \pi )-4
   K_x R_{\{1,-\}} \cos \left(\frac{P \pi }{2}\right), \\
  d &=&R_{\{2,+\}} \sin
   (P \pi ),\\
  f &=&-\sqrt{2} \left(R_{\{2,-\}} \cos (P \pi )+2
   K_x R_{\{1,+\}} \sin \left(\frac{P \pi }{2}\right)\right),\\
  g &=&\phantom{-}\sqrt{2} \left(
   R_{\{2,-\}} \cos (P \pi )-2
   K_x R_{\{1,+\}} \sin \left(\frac{P \pi }{2}\right)\right), 
\end{eqnarray}
with the functions
\begin{eqnarray}
  R_{\{1,\pm\}} &=& \frac{\left(1\pm(-1)^{n+n'}\right)(n^2+n'^2-P^2)}{((n-n')-P
   ) ((n+n')-P) ((n-n')+P)
   ((n+n')+P)} \label{R-function1},\\
  R_{\{2,\pm\}} &=& \frac{\left(1\pm(-1)^{n+n'}\right)(n^2+n'^2-
   (2P)^2)}{((n-n')-2 P) ((n+n')-2
   P) ((n-n')+2 P) ((n+n')+2
   P)}.\nonumber\\
   &&\label{R-function2}
\end{eqnarray}
\end{widetext}
\bibliographystyle{apsrev}
\bibliography{WenkKettemannV2.bib}
\end{document}